\documentclass[11pt]{article}
\usepackage{verbatim,amsmath,amssymb}
\usepackage{epsfig,float,color}
\usepackage{epstopdf}
\usepackage{geometry}
\usepackage{setspace}
\usepackage{wrapfig}
\usepackage{hyperref}
\usepackage[utf8]{inputenc}
\usepackage{graphicx}
\usepackage{flushend}
\usepackage[linesnumbered,ruled]{algorithm2e}

\usepackage[font={footnotesize}]{caption}
\usepackage[font={footnotesize}]{subcaption}
\usepackage{footnote}
\usepackage{multicol}
\usepackage{multirow}
\usepackage{enumitem}   
\usepackage{soul}
\usepackage{framed}
\usepackage[titletoc,title]{appendix}
\usepackage{bm}
\usepackage{siunitx}
\usepackage{cite}
\usepackage[makeroom]{cancel} 
\usepackage{natbib}
\usepackage{hyperref}
\definecolor{darkblue}{rgb}{0,0,1}
\hypersetup{pdftex=true, colorlinks=true, breaklinks=true, linkcolor=magenta, menucolor=magenta, citecolor=magenta, urlcolor=magenta}

\renewcommand{\bar}[1]{\tilde{#1}}

\newcommand{\tr}[1]{{#1}^{\!\top}}

\newcommand{\inv}[1]{{#1}^{\text{-}1}}

\newcommand{\der}[2]{\frac{d\!{#1}}{d\!{#2}}}
\newcommand{\pd}[2]{\frac{\partial\!{#1}}{\partial\!{#2}}}

\newenvironment{rcases}
{\left.\begin{aligned}}
	{\end{aligned}\right\rbrace}

\definecolor{darkblue}{rgb}{0,0,1}
\hypersetup{pdftex=true, colorlinks=true, breaklinks=true, linkcolor=darkblue, menucolor=darkblue, pagecolor=darkblue, citecolor=darkblue, urlcolor=darkblue}

\usepackage[dvipsnames]{xcolor}
\begin{document}
	
	\begin{center}
		\Large{\bf{Topology Optimization of Fluidic Pressure Loaded Structures and Compliant Mechanisms using the Darcy Method}}\\
		
	\end{center}
	
	\begin{center}
		\large{Prabhat Kumar$\footnote{Present Address: Department of Mechanical Engineering, Solid Mechanics, Technical University of Denmark, 2800 Kgs. Lyngby, Denmark}^{,}$ $\footnote{corresponding author: pkumar@mek.dtu.dk,\,prabhatkumar.rns@gmail.com}$, Jan S. Frouws, and Matthijs Langelaar}
		\vspace{4mm}
		
		\small{\textit{Department of Precision and Microsystems Engineering, Faculty of 3mE,
				Delft University of Technology, Mekelweg 2, 2628 CD, Delft, The Netherlands}}
		
		Published\footnote{This pdf is the personal version of an article whose final publication is available at \href{https://link.springer.com/article/10.1007/s00158-019-02442-0}{https://www.springer.com/journal/158}} 
		in \textit{Structural and Multidisciplinary Optimization}, 
		\href{https://doi.org/10.1007/s00158-019-02442-0}{DOI:10.1007/s00158-019-02442-0} \\
		Submitted on 24.~August 2019, Revised on 17. October 2019, Accepted on 22. October 2019
		
	\end{center}
	
	\vspace{3mm}
	\rule{\linewidth}{.15mm}
	{\bf Abstract:}
	In various applications, design problems involving structures and compliant mechanisms experience fluidic pressure loads. During topology optimization of such design problems, these loads adapt their direction and location with the evolution of the design, which poses various challenges.  A new density-based topology optimization approach using Darcy's law in conjunction with a drainage term is presented to provide a continuous and consistent treatment of design-dependent fluidic pressure loads. The porosity of each finite element and its drainage term are related to its density variable using a Heaviside function, yielding a smooth transition between the solid and void phases. A design-dependent pressure field is established using Darcy's law and the associated PDE is solved using the finite element method. Further, the obtained pressure field is used to determine the consistent nodal loads. The approach provides a computationally inexpensive evaluation of load sensitivities using the adjoint-variable method.  To show the efficacy and robustness of the proposed method, numerical examples related to fluidic pressure loaded stiff structures and small-deformation compliant mechanisms are solved. For the structures, compliance is minimized, whereas for the mechanisms a multi-criteria objective is minimized with given resource constraints.   \\
	

	\vspace{-4mm}
	\rule{\linewidth}{.15mm}
	
	\section{Introduction}\label{Intro}
In the last three decades, various topology optimization (TO) methods have been presented, and most have meanwhile attained a mature state. In addition, their popularity as design tools for achieving solutions to a wide variety of problems involving single/multi-physics is growing consistently. Among these, design problems involving fluidic pressure loads\footnote{Henceforth we write \textquotedblleft pressure loads" instead of \textquotedblleft fluidic pressure loads" throughout the manuscript for simplicity.} pose several unique challenges, e.g., (i) identifying the structural boundary to apply such loads, (ii) determining the relationship between the pressure loads and the design variables, i.e., defining a design-dependent and continuous pressure field, and (iii) efficient calculation of the pressure load sensitivities. Such problems can be encountered in various applications \citep{Hammer2000} such as air-, water- and/or snow-loaded civil and mechanical structures (aircraft, pumps, pressure containers, ships, turbomachinery), pneumatically or hydraulically actuated soft robotics or compliant mechanisms and pressure loaded mechanical metamaterials, e.g. \citep{zolfagharian2016evolution,yap2016high}, to name a few. Note, the shape or topology and  performance of the optimized structures or compliant mechanisms are directly related to the magnitude, location, and direction of the pressure loads which vary with the design. In this paper, a novel approach addressing the aforementioned challenges to optimize and design pressure loaded structures and mechanisms is presented. Hereby we target a density-based TO framework.

\begin{figure*}[h!]
	\begin{subfigure}[t]{0.50\textwidth}
		\centering
		\includegraphics[scale=1]{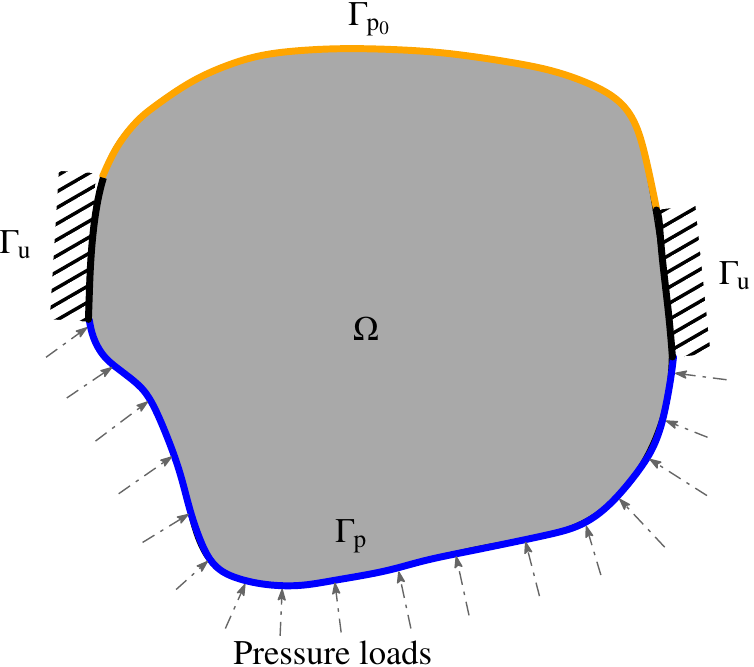}
		\caption{A design problem with pressure loading}
		\label{fig:sec1fig1a}
	\end{subfigure}
	~ \qquad
	\begin{subfigure}[t]{0.5\textwidth}
		\centering
		\includegraphics[scale=1]{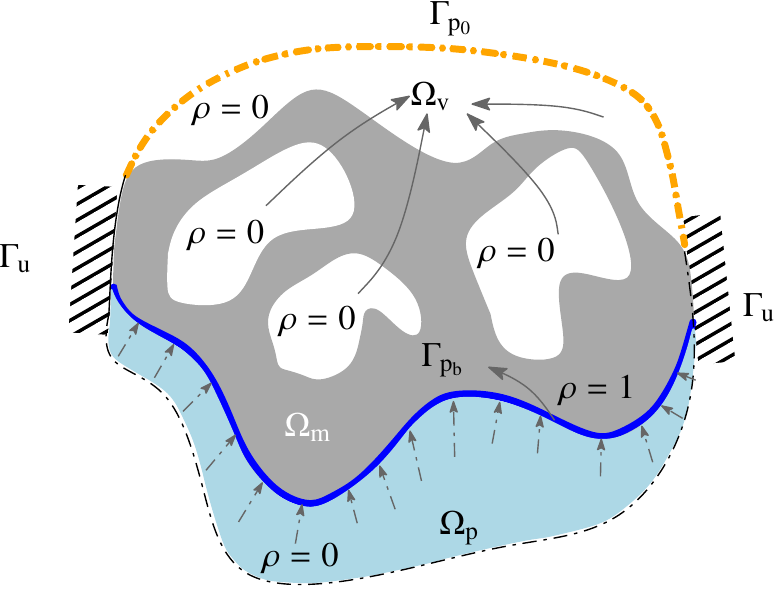}
		\caption{A representative solution to (a)}
		\label{fig:sec1fig1b}
	\end{subfigure}
	\caption{(\subref{fig:sec1fig1a}) A schematic diagram of a general design optimization problem experiencing pressure loading (depicted via dash-dotted arrows) on boundary $\mathrm{\Gamma}_\mathrm{p}$. (\subref{fig:sec1fig1b}) A representative solution to the problem in Fig.~(\subref{fig:sec1fig1a}). $\mathrm{\Omega},\,\mathrm{\Omega}_\mathrm{p}\, (\mathrm{\rho}=0),\,\mathrm{\Omega}_\mathrm{m}\,(\mathrm{\rho=1}),\,\text{and}\, \mathrm{\Omega}_\mathrm{v} \,(\mathrm{\rho}=0)$ indicate design domain, pressure (fluid) domain (void regions with pressurized boundary  $\mathrm{\Gamma}_\mathrm{p_b}$), mechanical design and void domain, respectively. \textbf{Key:} $\mathrm{\Gamma}_\mathrm{p_b}-$ evolving pressure boundary, $\mathrm{\Gamma}_\mathrm{p_0}-$ zero pressure boundary, $\mathrm{\Gamma}_\mathrm{u}-$ boundary with fixed displacements, $\rho-$ material density.}\label{fig:sec1fig1}
\end{figure*}
In line with the outlined applications, we are not only interested in optimizing \textit{pressure-loaded stiff structures}, but also in generating \textit{pressure-actuated compliant mechanisms (CMs)}. CMs are monolithic continua which transfer or transform energy, force or motion into desired work. Their performance relies on the motion obtained from the deformation of their flexible branches. The use of such mechanisms is on the rise in various applications as these mechanisms provide many advantages \citep{frecker1997topological} over their rigid-body counterparts. In addition, for a given input actuation, the output characteristic of a compliant mechanism can be customized, for instance, to achieve either output displacement in a certain desired fashion, e.g., path generation \citep{saxena2001topology,kumar2016synthesis}, shape morphing \citep{lu2003design} or maximum/minimum resulting (contact) force wherein grasping of an object is desired \citep{saxena2013contact}. \cite{Bendsoe2003} and \cite{deepak2009comparative} provide/mention various TO methods to synthesize structures and compliant mechanisms for the applications wherein input loads and constraints are considered invariant during the optimization. However, as mentioned above, a wide range of different applications with pressure loads can be found. 
 A schematic diagram for a general problem with pressure loads is depicted in Fig.~\ref{fig:sec1fig1a}, whereas Fig.~\ref{fig:sec1fig1b} is used to represent a schematic solution to the design problem with different optimized regions. A key problem characteristic is that the pressure-loaded surface is not defined \textit{a priori}, but that it can be modified by the optimization process (Fig.~\ref{fig:sec1fig1b}) to maximize actuation or stiffness. Below, we review the proposed TO methods that involve pressure-loaded boundaries, for either structures or mechanism designs.

\cite{Hammer2000} were first to present a TO method involving pressure loads. Thereafter, several approaches have been proposed to apply and provide a proper treatment of such loads in TO settings, which can be broadly classified into: (i) methods using boundary identification schemes \citep{Hammer2000,Du2004a,Zheng2009,Lee2012a,Fuchs2004,Li2018}, (ii) level set method based approaches \citep{gao2004topology,xia2015topology,Li2010}, and (iii) approaches involving special methods, i.e. which avoid detecting the loading surface \citep{Chen2001,bourdin2003design,Sigmund2007,Zhang2008,vasista2012design,Panganiban2010}. 

Boundary identification techniques, in general, are based on \textit{a priori} chosen threshold density  $\rho_T$, i.e., iso-density curves/surfaces are identified. \cite{Hammer2000} used the iso-density approach to identify the pressure loading facets $\mathrm{\Gamma}_\mathrm{p_b}$ (Fig.~\ref{fig:sec1fig1b}) which they further interpolated via B\'{e}zier spline curves to apply the pressure loading. However, as per \cite{Du2004a} this iso-density (isolines) method may furnish isoline-islands and/or separated isolines. Consequently, valid loading facets may not be achieved. In addition, this method requires predefined starting and ending points for $\mathrm{\Gamma_{p_b}}$ \citep{Hammer2000}. \cite{Du2004a} proposed a modified isolines technique to circumvent abnormalities associated with the isolines method. Refs. \citep{Hammer2000,Du2004a} evaluated the sensitivities of the pressure load with respect to design variables using an efficient finite difference formulation. \cite{Lee2012a}  presented a method wherein one does not need to define starting and ending points \textit{a priori}. In addition, they provided an analytical approach to calculate load sensitivities. Moreover, these studies \citep{Hammer2000,Du2004a,Lee2012a} considered sensitivities of the pressure loads, however they are confined to only those elements which are exposed to the pressure boundary loads $\mathrm{\mathrm{\Gamma_{p_b}}}$. 

  \cite{Fuchs2004} proposed a method wherein the evolving pressure loading boundary $\mathrm{\Gamma_{p_b}}$ is predefined using an additional set of variables, which are also optimized along with the design variables. \cite{Zhang2008} proposed an element-based search method to locate the load surface. They used the actual boundary of the finite elements (FEs) to construct the load surface and thereafter, transferred pressure to corresponding element nodes directly. \cite{Li2018} introduced an algorithm based on digital image processing and regional contour tracking to generate an appropriate pressure loading surface. They transferred pressure directly to nodes of the FEs. The methods presented in this paragraph do not account for load sensitivities within their TO setting.

As per \cite{Hammer2000}, if the evolving pressure loaded boundary $\mathrm{\Gamma_{p_b}}$ coincides with the edges of the FEs then the load sensitivities with respect to design variables vanish or can be disregarded. Consequently, $\mathrm{\Gamma_{p_b}}$ no longer remains sensitive to infinitesimal alterations in the design variables (density fields) unless the threshold value $\rho_T$ is passed and thus, $\mathrm{\Gamma_{p_b}}$ jumps directly to the edges of a next set of FEs in the following TO iteration. Note that load sensitivities however may critically affect the optimal material layout of a given design problem, especially those pertaining to  compliant mechanisms, as we will show in Sec. \ref{LS}. Therefore, considering load sensitivities in problems involving pressure loads is  highly desirable. In addition, ideally these sensitivities should be straightforward to compute, implement and computationally inexpensive. 

In contrast to density-based TO, in level-set-based approaches an implicit boundary description is available that can be used to define the pressure load. On the other hand, being based on boundary motion, level-set methods tend to be more dependent on the initial design \citep{van2013level}. \cite{gao2004topology} employed a level set function (LSF) to represent the structural topology and overcame difficulties associated with the description of boundary curves in an efficient and robust way. \cite{xia2015topology} employed two zero-level sets of two LSFs to  represent the free boundary and the pressure boundary separately. \cite{Wang2016} employed the Distance Regularized Level Set Evolution (DRLSE) \citep{Li2010} to locate the structural boundary. They used the zero level contour of an LSF to represent the loading boundary but did not regard load sensitivities. Recently, \cite{Picelli2019} proposed a method wherein Laplace's equation is employed to compute hydrostatic fluid pressure fields, in combination with interface tracking based on a flood fill procedure. Shape sensitivities in conjunction with Ersatz material interpolation approach are used within their approach.

Given the difficulties of identifying a discrete boundary within density-based TO and obtaining consistent sensitivity information, various  researchers have employed special/alternative methods (without identifying pressure loading surfaces directly) to design structures experiencing pressure loading. \cite{Chen2001} presented an approach based on applying a fictitious thermal loading to solve pressure loaded problems. \cite{Sigmund2007} employed a mixed displacement-pressure formulation based finite element method in association with three-phase material (fluid/void/solid). Therein, an extra (compressible) void phase is introduced in the given design problem while limiting the volume fraction of the fluid phase and also, the mixed finite element methods have to fulfill the BB-condition which guarantees the stability of the element formulation \citep{zienkiewicz2005finite}. \cite{bourdin2003design} also used three-phase material to solve such problems. \cite{Zheng2009} introduced a pseudo electric potential to model evolving structural boundaries. In their approach, pressure loads were directly applied upon the edges of FEs and thus, they did not account for load sensitivities. Additional physical fields or phases are typically introduced in these methods to handle the pressure loading. Our method follows a similar strategy based on Darcy's law, which has not been reported before.

This paper presents a new approach to design both structures and compliant mechanisms loaded by design-dependent pressure loads using density-based topology optimization. The presented approach uses Darcy's law in conjunction with a \textit{drainage} term (Sec. \ref{sec2.1.1}) and standard FEs,  for modeling and providing a suitable treatment of pressure loads. The drainage term is necessary to prevent pressure loads on structural boundaries that are not in contact with the pressure source, as explained in Sec. \ref{sec2.1.1}. Darcy's law is adapted herein in a manner that the porosity of the FEs can be taken as design (density) dependent (Sec. \ref{sec2.1}) using a smooth Heaviside function facilitating smoothness and differentiability.  Consequently, prescribed pressure loads are transferred into a design dependent pressure field using a PDE (Sec. \ref{sec2.2.1}) which is further solved using the finite element method. The determined pressure field is used to evaluate consistent nodal forces using the FE method (Sec. \ref{sec2.2.2}). This two step process offers a flexible and tunable method to apply the pressure loads and also, provides distributed load sensitivities, especially in the early stage of optimization. The latter is expected to enhance the exploratory characteristics of the TO process.

 In addition, regarding applications most research on topology optimization involving pressure loads has thus far focused on compliance minimization problems and, a thorough search yielded only two research articles for designing pressure-actuated compliant mechanisms. \cite{vasista2012design} employed the three-phase method proposed in \citep{Sigmund2007} to generate such mechanisms actuated via pressure loads whereas \cite{Panganiban2010} also used the three-phase method but in association with a displacement-based nonconforming FE method, which is not a standard FE approach. Herein, using the presented method, we not only design \textit{pressure-loaded structures} but also \textit{pressure-actuated compliant mechanisms}, which suggests the novel potentiality of the method. 
 
In summary, we present the following new aspects:

\begin{itemize}
	\item Darcy's law is used with a drainage term to identify evolving pressure loading boundary which is performed by solving an associated PDE, 	
	\item the approach facilitates computationally inexpensive evaluation of the load sensitivities with respect to design variables using the adjoint-variable method,
	\item the load sensitivities are derived analytically and consistently considered within the presented approach while synthesizing structures and compliant mechanisms experiencing pressure loading,
	\item the importance of load sensitivity contributions, especially in the case of compliant mechanisms, is demonstrated,
	\item the method avoids explicit description of the pressure loading boundary (which proves cumbersome to extend to 3D),
	\item  the robustness and efficacy of the approach is demonstrated via various standard design problems related to structures and compliant mechanisms,
	\item the method employs standard linear FEs, without the need for special FE formulations. 
\end{itemize}

The remainder of the paper is organized as follows: Sec. \ref{Sec2} describes the modeling of pressure loading via Darcy's law with a drainage term. Evaluation of consistent nodal forces from the obtained pressure field is presented therein. In Sec. \ref{Sec3}, the topology optimization problem formulation for pressure loaded structures and small-deformation compliant mechanisms is presented with the associated sensitivity analysis. In addition, the presented method is verified using a pressure-loaded structure problem on a coarse mesh. Sec. \ref{Sec4} presents the solution of various benchmark design problems involving pressure loaded structures and small deformation compliant mechanisms. Lastly, conclusions are drawn in Sec. \ref{Sec5}. 

	\section{Modeling of Design Dependent Loading}\label{Sec2}

The material boundary of a given design domain  $\mathrm{\Omega}$ evolves as the TO progresses while forming an optimum material layout. Therefore, it is challenging especially in the initial stage of the optimization to locate an appropriate loading boundary $\mathrm{\Gamma}_\mathrm{p_b}$ for applying the pressure loads. In addition, while designing especially pressure-actuated compliant mechanisms, establishing a design dependent and continuous pressure field would aid to TO. Herein, Darcy's law in conjunction with the drainage term, a volumetric material-dependent pressure loss, is employed to establish the pressure field as a function of material density vector $\bm{\rho}$.
\subsection{Darcy's law}\label{sec2.1}
 Darcy's law  defines the ability of a fluid to flow through  porous media such as rock, soil or sandstone. It states that fluid flow through a unit area is directly proportional to the pressure drop per unit length $\nabla p$ and inversely proportional to the resistance of the porous medium to the flow $\mu$ \citep{batchelor2000introduction}. Mathematically,

 \begin{equation}\label{sec2:eq1}
 \bm{q} = -\frac{\kappa}{\mu}\;\nabla p \quad = -K \;\nabla p,
 \end{equation}
 \begin{figure}[h!]
 	\centering
 	\includegraphics[scale = 1]{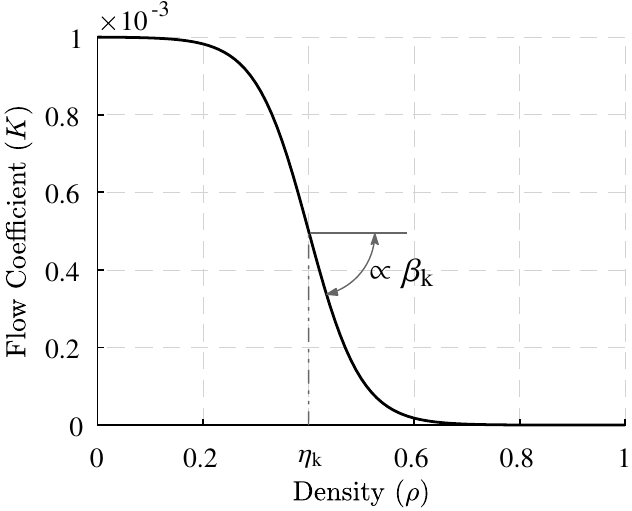}
 	\caption{A smooth Heaviside function is used to represent the density dependent flow coefficient $K(\rho_e)$. For the plot, $\eta_\mathrm{k }= 0.4$ and $\beta_\mathrm{k} = 10$ have been used. One notices that when $\eta_\mathrm{k}>\rho_e$,  $K(\rho_e) = k_\mathrm{v}$ and when $\eta_\mathrm{k}<\rho_e$, $K(\rho_e) = k_\mathrm{s}$.}
 	\label{fig:sec2fig1}
 \end{figure}
\begin{figure*}[h!]
	\begin{subfigure}[t]{0.30\textwidth}
		\centering
		\includegraphics[scale=0.75]{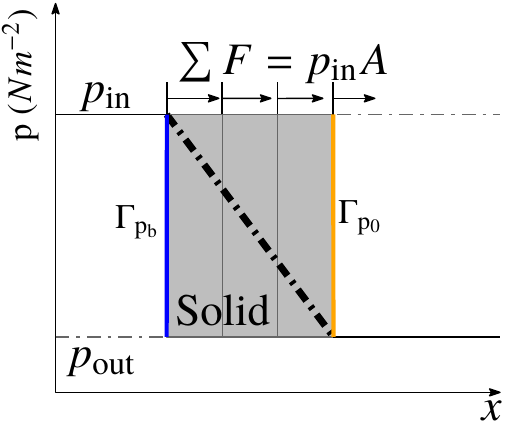}
		\caption{pressure drop over a single boundary without drainage term}
		\label{fig:sec2fig2a}
	\end{subfigure}
	\begin{subfigure}[t]{0.3\textwidth}
		\centering
		\includegraphics[scale=0.75]{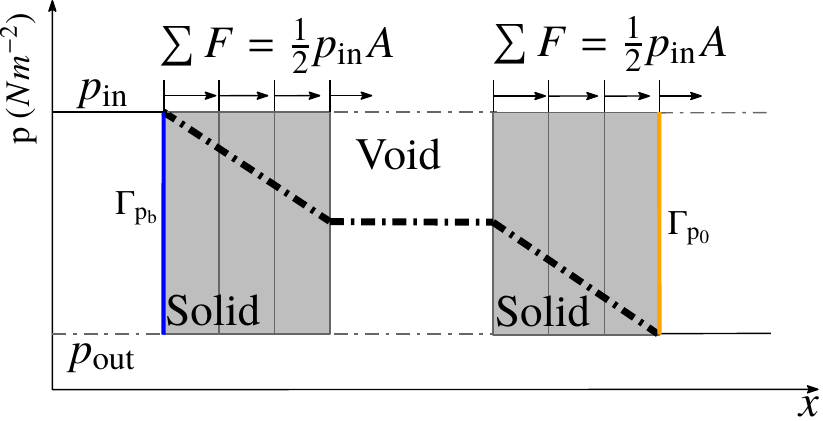}
		\caption{pressure drop over two boundaries without drainage term}
		\label{fig:sec2fig2b}
	\end{subfigure}
	~~ \qquad
	\begin{subfigure}[t]{0.3\textwidth}
		\centering
		\includegraphics[scale=0.75]{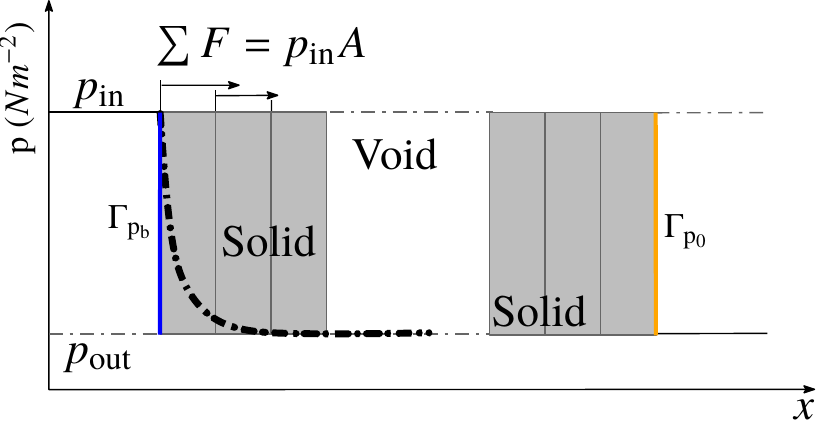}
		\caption{pressure drop over two boundaries with drainage term}
		\label{fig:sec2fig2c}
	\end{subfigure}
	\caption{Behaviour of a 1-D pressure field (thick dash-dotted lines/curves) when using Darcy's law with porous material (with 3FEs). (\subref{fig:sec2fig2a}) pressure drop  over a single wall  (\subref{fig:sec2fig2b}) undesirable condition wherein pressure drop takes place over multiple walls. When an additional drainage term, i.e. a volumetric density-dependent pressure loss, is considered then the pressure drop over multiple walls takes the form shown in (\subref{fig:sec2fig2c}). This is the desired behaviour for a TO setting. $A$ is the cross section area of the porous medium used in this 1D example.}\label{fig:sec2fig2}
\end{figure*}
 where 	$\bm{q}, \,\kappa, \,\mu,\,\text{and},\,\nabla p$ represent the flux ($\si{\meter\per\second}$), permeability ($\si{\meter\squared}$), fluid viscosity ($\si{\newton\per\square\meter\second}$) and pressure gradient ($\si{\newton\per\cubic\meter}$), respectively. Further, $K$ ($\si{\meter\tothe{4}\per\newton\per\second}$) is termed herein as a flow coefficient\footnote{$K = \frac{\kappa}{\mu}$ is termed `flow coefficient' herein, noting the fact that this terminology is  however sometimes used in literature with a different meaning.} which expresses the ability of a fluid to flow through a porous medium. The flow coefficient of each FE is assumed to be related to element density $\rho_e$. In order to differentiate between void ($\rho_e=0$) and solid  ($\rho_e=1$)  states of a FE, and at the same time ensuring a smooth and differentiable transition, $K(\rho_e)$ is modeled using a smooth Heaviside function as:
 \begin{equation}\label{sec2:eq2}
 K(\rho_e) = k_\mathrm{v} - k_\mathrm{vs} \frac{\tanh{\left(\beta_\mathrm{k}\eta_\mathrm{k}\right)}+\tanh{\left(\beta_\mathrm{k}(\rho_e - \eta_\mathrm{k})\right)}}{\tanh{\left(\beta_\mathrm{k} \eta_\mathrm{k}\right)}+\tanh{\left(\beta_\mathrm{k}(1 - \eta_\mathrm{k})\right)}},
 \end{equation}
where $k_\mathrm{vs}= (k_\mathrm{v}-k_\mathrm{s})$, $k_\mathrm{v}$ and $k_\mathrm{s}$ are the flow coefficients for a void and solid FE, respectively. Further, $\eta_\mathrm{k}$ and $\beta_\mathrm{k}$ are two adjustable parameters which control the position of the step and the slope, respectively (Fig.~\ref{fig:sec2fig1}). For sufficiently high $\beta_\mathrm{k}$, when $\eta_\mathrm{k}>\rho_e$,  $K(\rho_e) = k_\mathrm{v}$ while when $\eta_\mathrm{k}<\rho_e$, $K(\rho_e) = k_\mathrm{s}$. In view of the permeability of an impervious material and viscosity of air, the flow coefficient of a solid element is chosen to be $k_\mathrm{s} = \SI{e-10}{\meter\tothe{4}\per\newton\per\second}$, whereas, $k_\mathrm{v} = \SI{e-3}{\meter\tothe{4}\per\newton\per\second}$ is taken to mimic a free flow with low resistance through the void regions.

Our intent is to smoothly and continuously distribute the pressure drop over a certain penetration depth of the solid facing the pressure source. To examine the interaction between structural features and applied pressure under Darcy's law, consider Fig.~\ref{fig:sec2fig2a}. Darcy's law renders a gradual pressure drop from the inner pressure boundary $\mathrm{\Gamma_{p_b}}$ to the outer pressure boundary $\mathrm{\Gamma_{p_0}}$ (Fig.~\ref{fig:sec2fig2a}). Consequently, equivalent nodal forces appear within the material as well as upon the associated boundaries. This penetrating pressure, originating because of Darcy's law, is a smeared-out version of an applied pressure load on a sharp boundary or interface\footnote{used in the approaches based on boundary identification}. Note that, summing up the contributions of penetrating loads gives the resultant load. It is assumed that local differences in the load application have no significant effect on the global behaviour of the structure, in line with the Saint-Venant principle. The validity of this assumption will be checked later in a numerical example (Sec. \ref{sec3.5}).

\subsubsection{Drainage term}\label{sec2.1.1}
 Application of Darcy's law alone introduces an undesired pressure distribution in the model when multiple walls are encountered between $\mathrm{\Gamma_{p_b}} (p_\mathrm{in})$ and $\mathrm{\Gamma_{p_0}}(p_\mathrm{out})$. That is, the pressure does not completely drop over the first boundary as illustrated in Fig.~\ref{fig:sec2fig2b}. To mitigate this issue, we introduce a drainage term, which is a volumetric density-dependent pressure loss, as
\begin{equation}\label{sec2:eq3}
 {Q}_\mathrm{drain} = - H(\rho_e) (p - p_{\mathrm{out}}),
\end{equation}
 where ${Q}_\mathrm{drain}$ denotes volumetric drainage per second in a unit volume ($\si{\per\second}$). $H,\,p,\,p_\mathrm{out}$ are drainage coefficient ($\si{\meter\tothe{2}\per\newton\per\second}$), continuous pressure field ($\si{\newton\per\square\meter}$), external pressure\footnote{in this work $p_{\mathrm{out}}=0$} ($\si{\newton\per\square\meter}$), respectively. Conceptually, this term should drain/absorb the flow in the exterior structural boundary layer exposed to the pressure source, so that negligible flow (and pressure) acts on interior structural boundaries. 
 
\begin{figure}[h!]
 	\centering
 	\includegraphics[scale = 1]{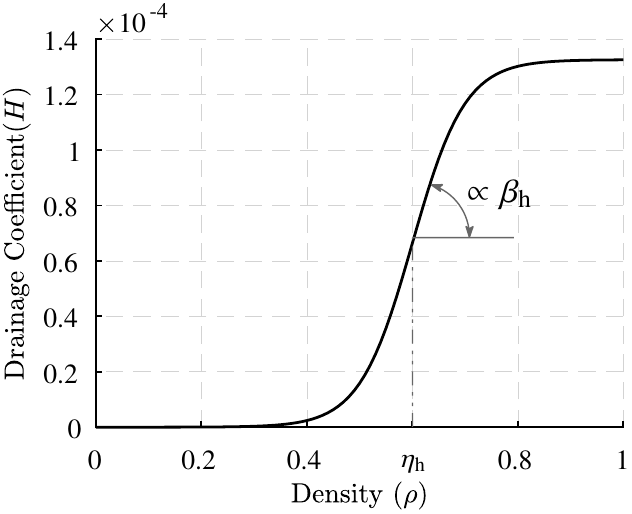}
 	\caption{A Heaviside function is used to represent the drainage coefficient $H (\rho_e)$ using the Heaviside parameters $\eta_\mathrm{h} =0.6$ and $\beta_\mathrm{h} =10$. Herein, $r = 0.1$, $\Delta s =\SI{2}{mm}$ and $k_\mathrm{s} = \SI{e-10}{\meter\tothe{4}\per\newton\per\second}$ are considered  to find $h_s$ in Eq.~(\ref{sec2:eq5}), which is used in Eq.~(\ref{sec2:eq4}) for  evaluating $H (\rho_e)$. It can be seen that when $\eta_\mathrm{h}>\rho_e$,  $H(\rho_e)\to 0$ and when $\eta_\mathrm{h}<\rho_e$, $H(\rho_e)\to h_\mathrm{s}$.}
 	\label{fig:sec2fig3}
 \end{figure}

 Similar to flow coefficient $K (\rho_e)$, the drainage coefficient $H(\rho_e)$ is also modeled using a smooth Heaviside function such that pressure drops to zero when $\rho_e = 1$ (Fig.~\ref{fig:sec2fig2c}). It is given by:
 
 \begin{equation}\label{sec2:eq4}
 H(\rho_e)    = h_{\mathrm{s}} \frac{\tanh{\left(\beta_\mathrm{h} \eta_\mathrm{h}\right)} + \tanh{\left(\beta_\mathrm{h} (\rho_e-\eta_\mathrm{h})\right)}}{\tanh{\left(\beta_\mathrm{h} \eta_\mathrm{h}\right)} + \tanh{\left(\beta_\mathrm{h} (1-\eta_\mathrm{h})\right)}},
 \end{equation}
 where, $\beta_\mathrm{h}$ and $\eta_\mathrm{h}$ are adjustable parameters similar to $\beta_\mathrm{k}$ and $\eta_\mathrm{k}$. $h_\mathrm{s}$ is the drainage coefficient of solid, which is used to control the thickness of the pressure-penetration layer. This formulation can effectively control the location and depth of penetration of the applied pressure. Note, $h_\mathrm{s}$ is related to $k_\mathrm{s}$ (Appendix \ref{appendA}) as:
 \begin{equation}\label{sec2:eq5}
 h_\mathrm{s} =\left(\frac{\ln{r}}{\Delta s}\right)^2 k_\mathrm{s},
 \end{equation} 
 where $r$ is the ratio of input pressure at depth $\Delta$s, i.e., $p|_{\Delta s} = rp_\mathrm{in}$. Further, $\Delta s$ is the penetration depth of pressure, which can be set to the width or height of few FEs. Fig.~\ref{fig:sec2fig3} depicts a plot for the drainage coefficient $ H(\rho_e)$ as a function of density. Note that the Heaviside parameters used in this plot are the same as those employed in Fig.~\ref{fig:sec2fig1}. 
\subsection{Finite Element Formulation}\label{sec2.2}
This section presents the FE formulation of the proposed pressure load based on Darcy's law, wherein the approach employs the standard FE method \citep{zienkiewicz2005finite} to solve the associated boundary value problems to determine the pressure and displacement fields. Standard 2D quadrilateral elements with bilinear shape functions are employed to parameterize the design domain. First, in addition to the Darcy equation (Eq.~\ref{sec2:eq1}), the equation of state using the law of conservation of mass in view of incompressible fluid is derived. Thereafter, the consistent nodal loads are determined from the derived pressure field. 
 \subsubsection{State Equation}\label{sec2.2.1}
\begin{figure}[h!]
	\centering
	\includegraphics[scale = 1]{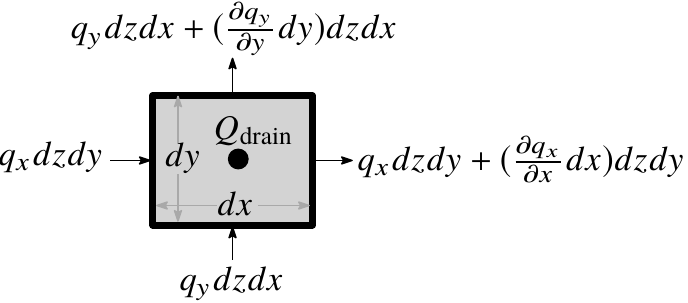}
	\caption{In- and outflow of an infinitesimal element with volume, $d V = d x d y d z$. $Q_\mathrm{drain}$ is the volumetric drainage per second in a unit volume.}
	\label{fig:sec2fig4}
\end{figure}

  Fig.~\ref{fig:sec2fig4} shows in- and outflow  through an infinitesimal volume element $\mathrm{\Omega_e}$. Now, using the conservation of mass for incompressible fluid one writes:
\begin{equation} \label{sec2:eq6}
\begin{aligned}
\left(q_{x}d y\;+\; q_{y}dx \;+\; {Q}_\mathrm{drain}d xd y \right)d z =&\\
\left(q_{x}d y \;+\; q_{y} d x \;+\; \left(\frac{\partial q_x}{\partial x}d x\right) d y \;+\;  \left(\frac{\partial q_y}{\partial y}d y\right) d x\right)d z, \\ \text{or,}\,\,
\frac{\partial q_x}{\partial x} + \frac{\partial q_y}{\partial y}-{Q}_\mathrm{drain}  = &0,\\ \text{or,}\,\,
\nabla\cdot\bm{q} -{Q}_\mathrm{drain} = &0.
\end{aligned}
\end{equation}
where $q_x$ and $q_y$ are the flux in $x$- and $y$-directions, respectively. In view of Eq.~(\ref{sec2:eq1}), Eq.~(\ref{sec2:eq6}) becomes:
\begin{equation}\label{sec2:eq7}
\nabla\cdot \left(K\nabla p(\bm{x})\right) + {Q}_\mathrm{drain} = 0.
\end{equation}
Now, for the finite element formulation, we use the Galerkin approach to seek an approximate solution $p (\bm{x})$ such that:
\begin{equation}\label{sec2:eq8}
\sum_{e=1}^{n_\mathrm{elem}}\left(\int_{\mathrm{\Omega}_e}\nabla\cdot \left(K\nabla p(\bm{x})\right)w(\bm{x}) d V + \int_{\mathrm{\Omega}_e}{Q}_\mathrm{drain}w(\bm{x}) d V\right) = 0,
\end{equation}
for every $w (\bm{x})$ constructed from the same basis functions as those employed for $p(\bm{x})$. The total number of elements is indicated via $n_\mathrm{elem}$. In the discrete setting, within each $\mathrm{\Omega}_e{|_{e=1,\,2,\,3,\,\cdots,\,n_\mathrm{elem}}}$, we have
\begin{equation}\label{sec2:eq9}
p_e = \mathbf{N}_\text{p}\mathbf{p}_{e}, \qquad w = \mathbf{N}_\text{p} \mathrm{\bm{w}}_{e},
\end{equation}
where $\mathbf{N}_\text{p} = [N_1,\, N_2,\,N_3,\,N_4]$ are the bilinear shape functions in a physical element and $\mathbf{p}_{e} = \tr{[p_1,\,p_2,\,p_3,\,p_4]}$ is the nodal pressure. Now, with integration by parts and Greens' theorem,  Eq.~(\ref{sec2:eq8}) becomes on elemental level:
 
\begin{equation}\label{sec2:eq10}
\begin{aligned}
\int_{\mathrm{\Omega}_e} K\left(\nabla w (\bm{x})\right)\cdot \left(\nabla p(\bm{x})\right)d V + \int_{\mathrm{\Omega}_e}{Q}_\mathrm{drain}w(\bm{x}) d V \\= -\int_{\mathrm{\Gamma}_e} w(\bm{x})\bm{q}_\mathrm{\Gamma}.\bm{n}_edA,
\end{aligned}
\end{equation}
where $\bm{n}_e$ is the boundary normal on surface $\mathrm{\Gamma}_e$ and therein, $\bm{q}$ changes to $\bm{q}_\mathrm{\Gamma}$. In view of Eq.~(\ref{sec2:eq3}) and Eq.~(\ref{sec2:eq9}), Eq.~(\ref{sec2:eq10}) gives:

\begin{equation} \label{sec2:eq11}
\begin{aligned}
\underbrace{\int_{\mathrm{\Omega}_e}\left( K~ \tr{\mathbf{B}}_\mathrm{p} \mathbf{B}_\mathrm{p}   + H ~\tr{\mathbf{N}}_\mathrm{p} \mathbf{N}_\mathrm{p} \right)d V}_{\mathbf{A}_e}~\mathbf{p}_e=\\
\underbrace{\int_{\mathrm{\Omega}_e}~H~\tr{\mathbf{N}}_\mathrm{p} p_\mathrm{out} ~~d V -
	\int_{\mathrm{\Gamma}_e}~ \tr{\mathbf{N}}_\mathrm{p} \bm{q}_\mathrm{\Gamma} \cdot \bm{n}_e~~d A}_{\mathbf{f}_e},
\end{aligned}
\end{equation}
where $\mathbf{B}_\mathrm{p} =\nabla\mathbf{N}_\mathrm{p}$ and $\bm{q}_\mathrm{\Gamma}$ is the Darcy flux through the boundary $\mathrm{\Gamma}_e$. In global sense, i.e., after assembly , Eq.~(\ref{sec2:eq11}) is written as
\begin{equation}\label{sec2:eq12}
\mathbf{A}\mathbf{p} = \mathbf{f},
\end{equation}
where $\mathbf{A}$ is termed the global flow matrix, $\mathbf{p}$ and $\mathbf{f}$ are the global pressure vector and loading vector, respectively. Note, when $p_\mathrm{out}= 0$ and $q_\mathrm{\Gamma}=0$ then conveniently $\mathbf{f} = 0$ and therefore, the right hand side only contains the contribution from the prescribed pressure, which is the case we have considered while solving design problems in this paper.

\subsubsection{Pressure field to consistent nodal loads}\label{sec2.2.2}
\begin{figure}[h!]
	\centering
	\includegraphics[scale = 1]{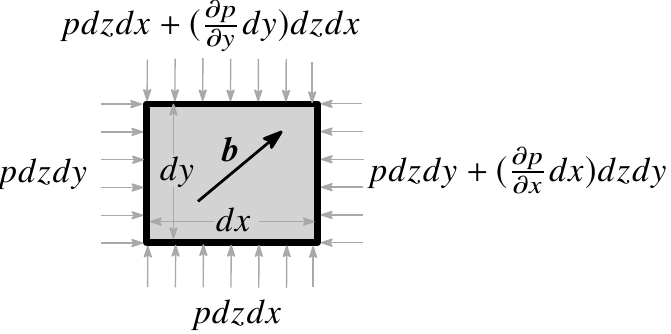}
	\caption{An infinitesimal element with volume, $d V = d x d y d z$. The pressure loads are shown using uniformly placed arrows on the boundary, are in equilibrium with the body force $\bm{b}$.}
	\label{fig:sec2fig5}
\end{figure}

The force resulting from the pressure field is expressed as an equivalent body force.  Fig.~\ref{fig:sec2fig5} depicts an infinitesimal volume element with  pressure loads acting on it, which is used to relate the pressure field ${p}(\bm{x})$ and body force $\bm{b}$.

Writing the force equilibrium equations, one obtains:

\begin{equation}\label{sec2:eq13}
\begin{bmatrix} 
pd z d y - pd z d y - \left(\pd{p}{x}d x\right)d z d y\\
pd z d x - pd z d x - \left(\pd{p}{y}d y\right)d z d x\\
pd x d y - pd x d y - \left(\pd{p}{z}d z\right)d x d y
\end{bmatrix} = \begin{bmatrix} b_x \\ b_y\\ b_z \end{bmatrix}d V,
\end{equation}
where, $b_x,\,b_y,\,\text{and}\,b_z$ are the components of the body force in $x,\,y,\,\text{and}\,z$ directions respectively. Eq.~(\ref{sec2:eq13}) can be written as\footnote{In 2D case, $d z$ is the thickness $t$ and $\pd{p}{z}=0$}:
\begin{equation}\label{sec2:eq14}
\quad \bm{b} dV = -\nabla p dV.
\end{equation}

In the discretized setting, $-\nabla p dV = -\mathbf{B}_\mathrm{p} \mathbf{p}_e dV$.  In general, the external elemental force originating from the body force $\bm{b}$ and traction $\bm{t}$ in a FE setting \citep{zienkiewicz2005finite}, can be written as:
\begin{equation}\label{sec2:eq15}
\mathbf{F}^e = \int_{\mathrm{\Gamma}_e} \tr{\mathbf{N}}_\mathbf{u}  \bm{t} \;dA \;+\int_{\mathrm{\Omega}_e} \tr{\mathbf{N}}_\mathbf{u} \bm{b} \;dV,
\end{equation}
where $\mathbf{N}_\mathbf{u} = [N_1\mathbf{I},\, N_2\mathbf{I},\,N_3\mathbf{I},\,N_4\mathbf{I}]$ with $\mathbf{I}$ as the identity matrix in $\mathcal{R}^2$ herein. In this work, we consider $\bm{t} = 0$. Thus,  Eq.~(\ref{sec2:eq15}) gives the consistent nodal loads on elemental level as:
\begin{equation}\label{sec2:eq16}
\mathbf{F}^e = - \int_{\mathrm{\Omega}_e} \tr{\mathbf{N}}_\mathbf{u} \nabla p d V = - \underbrace{\int_{\mathrm{\Omega}_e} \tr{\mathbf{N}}_\mathbf{u} \mathbf{B}_\mathrm{p}  d V}_{\mathbf{H}_e} \mathbf{p}_e.
\end{equation}
Next, in the global form, the consistent nodal loads $\mathbf{F}$ can be evaluated from the global pressure vector $\mathbf{p}$ (Eq.~\ref{sec2:eq12}) using the global conversion matrix $\mathbf{H}$ obtained by assembling all  such $\mathbf{H}_e$ as:
\begin{equation}\label{sec2:eq17}
\mathbf{F} = -\mathbf{H}\mathbf{p}.
\end{equation}
Note that H is independent of the design, the design-dependence of the loading enters through the pressure field obtained through Darcy's law (Eq.~\ref{sec2:eq12}).

	\section{Problem Formulation}\label{Sec3}
We follow the classical density-based TO formulation and employ the modified SIMP (Solid Isotropic Material and Penalization) approach \citep{sigmund2007morphology} to relate the element stiffness matrix of each element to its design variable. This is  realized by defining the Young's modulus of an element as:
\begin{equation}\label{sec3:eq1}
E_e(\rho_e) = E_\mathrm{min} + \rho_e^\zeta (E_0 - E_\mathrm{min}), \qquad \rho_e\in[0, \, 1]
\end{equation}
where, $E_0$ is the Young's modulus of the actual material, $E_\mathrm{min}$ is a significantly small Young's modulus assigned to the void regions, preventing the stiffness matrix from becoming singular, and $\zeta$ is a penalization parameter (generally, $\zeta =3$) which steers the TO towards \textquotedblleft 0-1\textquotedblright solutions. In the following subsections, we present the optimization problem formulations for the structures and CMs, discuss the sensitivity analysis for both type of problems and present a numerical verification study of the proposed Darcy-based pressure load formulation.
\subsection{Stiff structures}\label{sec3.1}
The standard formulation, i.e., minimization of compliance or strain energy is considered to design pressure loaded stiff structures \citep{Bendsoe2003} wherein the optimization problem is formulated as:

\begin{equation}\label{sec3:eq2}
\begin{rcases}
& \underset{\bm{\rho}}{\text{min}}
& &f_0^\mathrm{s}(\mathbf{u},\,\bm{\rho}) = \mathbf{\tr{u} K u} =  2SE\\
& \text{such that} & & (i)\,\, \mathbf{Ap} = \mathbf{0} \\
&  &&(ii)\,\,\mathbf{Ku = F} = -\mathbf{H p}\\
&  && (iii)\,\,\frac{ V(\bm{\rho})}{V^*}\le 1\\
&  && \mathbf{0}\le\bm{\rho}\le \mathbf{1}
\end{rcases},
\end{equation}
where $f_0^\mathrm{s}(\mathbf{u},\,\bm{\rho})$ is the compliance of the structure, $\mathbf{K}$ and $\mathbf{u}$ are the global stiffness matrix and displacement vector, respectively. $\mathbf{A}$, $\mathbf{H}$, $\mathbf{F}$ and $\mathbf{p}$ are the global flow matrix, conversion matrix,  nodal force vector and pressure vector, respectively. Further, $V(\bm{\rho})$ and ${V^*}$ are the material volume and the upper bound of volume respectively. Note, all mechanical equilibrium equations are satisfied under small deformation assumption. A standard nested optimization strategy is employed, wherein the boundary value problems $(i)$ and $(ii)$ (Eq.~\ref{sec3:eq2}) are solved in each iteration in combination with the respective boundary conditions.
\subsection{Compliant Mechanisms}\label{sec3.2}

In general, while designing compliant mechanisms, an objective stemming from a stiffness measure (e.g., compliance, strain energy) and a flexibility measure (e.g. output deformation) of the mechanisms is formulated and optimized \citep{saxena2000optimal}. The former measure provides adequate stiffness under the actuating loads while the latter one helps achieve the desired deformation at the output port. Note, a spring with certain stiffness $k_\mathrm{ss}$ representing the workpiece stiffness, is added at the output location. The spring motivates the optimization process to connect sufficient material to the output port/location. 
 
The flexibility-stiffness based multi-criteria formulation \citep{frecker1997topological,saxena2000optimal} is employed herein to design CMs. The proposed Darcy-based pressure load formulation is also expected to work with other CM formulations (\cite{deepak2009comparative}) with required modification e.g. \cite{Panganiban2010} to render suitable treatment for pressure loading cases, however this aspect has not been studied and is considered beyond the scope of this paper. As per \cite{saxena2000optimal}, the output deformation, measured in terms of mutual strain energy $(MSE)$, is maximized and the stored internal energy ($SE$) is minimized.  The optimization problem can be expressed as: 
 
 \begin{equation}\label{sec3:eq3}
 \begin{rcases}
 & \underset{\bm{\rho}}{\text{min}}
 & &{f_0^\mathrm{CM}}(\mathbf{u},\,\mathbf{v},\,\bm{\rho}) = -\frac{MSE (\mathbf{u},\,\mathbf{v},\,\bm{\rho})}{2SE(\mathbf{u},\,\bm{\rho})}\\
 & \text{such that} & & (i)\,\, \mathbf{Ap} = \mathbf{0 }\\
 &  &&(ii)\,\,\mathbf{Ku = F} = -\mathbf{H p}\\
  &  &&(iii)\,\,\mathbf{Kv = F_\mathrm{d}}\\
 &  && (iv)\,\,\frac{ V(\bm{\rho})}{V^*}\le 1\\
 &  && \mathbf{0}\le\bm{\rho}\le \mathbf{1}
 \end{rcases},
 \end{equation}
 where $f_0^\mathrm{CM}$ is the multi-criteria objective and $MSE = \tr{\mathbf{v}}\mathbf{Ku}$. Further,  $\mathbf{F}_\mathrm{d}$, the unit dummy force vector having the same direction as that of the output deformation, is used to evaluate $\mathbf{v}$ using $(iii)$ (Eq.~\ref{sec3:eq3}).    Other variables have the same definition as defined in Sec. \ref{sec3.1}.
 \subsection{Sensitivity Analysis}\label{sec3.3}
 In a gradient-based topology optimization, it is essential to determine sensitivities of the objective function and the constraints with respect to the design variables. In general, the formulated objective function depends upon both the state variables\footnote{In case of CM, state variables are $\mathbf{u}$ and $\mathbf{v}$ originated from input load and dummy load at output port, respectively.} $\mathbf{u}$,  solution to the mechanical equilibrium equations, and the design variables, the densities $\bm{\rho}$. The presented Darcy-based TO method facilitates use of the adjoint-variable approach to determine the sensitivity wherein an augmented performance function $\mathrm{\Phi} (\mathbf{u},\mathbf{v},\,\bm{\rho})$ can be defined using the objective function and the mechanical state equations as\footnote{Herein, a generic case of CM is considered.}:
 \begin{equation}\label{sec3:eq4}
\begin{aligned}
 {\mathrm{\Phi} (\mathbf{u},\mathbf{v},\,\bm{\rho})} = f_0(\mathbf{u},\mathbf{v},\bm{\rho}) + \tr{\bm{\lambda}}_1 \left(\mathbf{Ku +{H p}}\right) \\+ \tr{\bm{\lambda}}_2 ( \mathbf{Ap}) + \tr{\bm{\lambda}}_3 (\mathbf{Kv-F_\mathrm{d}}).
\end{aligned}
 \end{equation}
 The sensitivities are evaluated by differentiating Eq.~(\ref{sec3:eq4}) with respect to the design vector as: 
  \begin{equation}\label{sec3:eq5}
 \begin{aligned}
 \frac{d\mathrm{\Phi}}{d\bm{\rho}} =&
\underbrace{\left(\pd{f_0}{\mathbf{u}} + \tr{\bm{\lambda}}_1 \mathbf{K}\right)}_{\text{Term}\,1}\pd{\mathbf{u}}{\bm{\rho}} + \pd{f_0}{\bm{\rho}} + \tr{\bm{\lambda}}_1\pd{\mathbf{K}}{\bm{\rho}}\mathbf{u}
 \\+ &\underbrace{\left(\tr{\bm{\lambda}}_1 \mathbf{H} + \tr{\bm{\lambda}}_2 \mathbf{A}\right) }_{\text{{Term}} \,2}\pd{\mathbf{p}}{\bm{\rho}}+\tr{\bm{\lambda}}_2\pd{\mathbf{A}}{\bm{\rho}}\mathbf{p} \\
 +&\underbrace{\left(\pd{f_0}{\mathbf{v}} + \tr{\bm{\lambda}}_3 \mathbf{K}\right)}_{\text{Term}\,3}\pd{\mathbf{v}}{\bm{\rho}} + \tr{\bm{\lambda}}_3\pd{\mathbf{K}}{\bm{\rho}}\mathbf{v},
 \end{aligned}
 \end{equation} 
where $\bm{\lambda}_1,\,\bm{\lambda}_2$ and $\bm{\lambda_3}$ are the Lagrange multiplier vectors which are selected such that Term 1, Term 2 and Term 3 in Eq.~(\ref{sec3:eq5}) vanish, i.e.,
\begin{equation}\label{sec3:eq6}
\begin{rcases}
\tr{\bm{\lambda}}_1  &= -\pd{f_0(\mathbf{u},\, \mathbf{v},\,\bm{\rho})}{\mathbf{u}} \inv{\mathbf{K}}\\
\tr{\bm{\lambda}}_2  & = -\tr{\bm{\lambda}}_1 \mathbf{H}\inv{\mathbf{A}}\\
\tr{\bm{\lambda}}_3  &= -\pd{f_0(\mathbf{u},\, \mathbf{v},\,\bm{\rho})}{\mathbf{v}} \inv{\mathbf{K}}
\end{rcases}.
\end{equation} 	
Note, the evaluation of $\bm{\lambda}_2$ is nontrivial as degrees of freedom of both the displacement and pressure field are involved. Details of the evaluation of the multipliers are provided in Appendix~\ref{appendA2}. Now, Eq.~(\ref{sec3:eq6}) can be used in  Eq.~(\ref{sec3:eq5}) to determine the sensitivities as:
\begin{equation}\label{sec3:eq7}
\frac{d f_0}{d\bm{\rho}} = \pd{f_0}{\bm{\rho}} + \tr{\bm{\lambda}}_1\pd{\mathbf{K}}{\bm{\rho}}\mathbf{u} + \tr{\bm{\lambda}}_2\pd{\mathbf{A}}{\bm{\rho}}\mathbf{p} +  \tr{\bm{\lambda}}_3\pd{\mathbf{K}}{\bm{\rho}}\mathbf{v}.
\end{equation}

Note that vector $\mathbf{p}$ also includes the prescribed boundary pressures.
\subsubsection{Case I: Designing Structures}\label{sec3.3.1}
While designing structures,  the state variable $\mathbf{v}$ does not exist. In that case, one only needs to evaluate $\bm{\lambda}_1$ and $\bm{\lambda}_2$ herein to determine the sensitivities. Now, using Eq.~(\ref{sec3:eq2}) and Eq.~(\ref{sec3:eq6}) in Eq.~(\ref{sec3:eq7}) gives:
 \begin{equation}\label{sec3:eq8}
 \frac{d f_0^\mathrm{s}}{d\bm{\rho}} =  -\tr{\mathbf{u}}\pd{\mathbf{K}}{\bm{\rho}}\mathbf{u} + \underbrace{2\tr{\mathbf{u}} \mathbf{H}\inv{\mathbf{A}}\pd{\mathbf{A}}{\bm{\rho}}\mathbf{p}}_{\text{Load sensitivities}}.
 \end{equation}
The partial density derivative terms follow directly from the interpolations defined earlier.  

\subsubsection{Case II: Designing Compliant Mechanisms}\label{sec3.3.2}

To design CMs, all three adjoint variables $\bm{\lambda}_1,\,\bm{\lambda}_2$ and $\bm{\lambda}_3$ are needed to determine the sensitivities. Considering the objective function (Eq.~\ref{sec3:eq3}), Eq.~(\ref{sec3:eq6}) yields:
\begin{equation}\label{sec3:eq9}
\begin{rcases}
\tr{\bm{\lambda}}_1 &= \left(\frac{1}{2SE}\tr{\mathbf{v}} - \frac{MSE}{(2SE)^2}\tr{2\mathbf{u}}\right)\\
\tr{\bm{\lambda}}_2 &= -\left(\frac{1}{2SE}\tr{\mathbf{v}} - \frac{MSE}{(2SE)^2}\tr{2\mathbf{u}}\right) \mathbf{H}\inv{\mathbf{A}}\\
\tr{\bm{\lambda}}_3 & = \frac{1}{2SE}\tr{\mathbf{u}}
\end{rcases}.
\end{equation}
Now, in view of Eq.~(\ref{sec3:eq9}), the sensitivities can be evaluated as:
\begin{equation}\label{sec3:eq10}
\begin{aligned}
\frac{d f_0^\mathrm{CM}}{d\bm{\rho}} =  &\frac{MSE}{(2SE)^2}\left(-\tr{\mathbf{u}}\pd{\mathbf{K}}{\bm{\rho}}\mathbf{u}\right) + \frac{1}{2SE}\left(\tr{\mathbf{u}}\pd{\mathbf{K}}{\bm{\rho}}\mathbf{v}\right)+\\
&\underbrace{\frac{MSE}{(2SE)^2}\left(2\tr{\mathbf{u}} \mathbf{H}\inv{\mathbf{A}}\pd{\mathbf{A}}{\bm{\rho}}\mathbf{p}\right) + \frac{1}{2SE}\left(-\tr{\mathbf{v}} \mathbf{H}\inv{\mathbf{A}}\pd{\mathbf{A}}{\bm{\rho}}\mathbf{p}\right)}_{\text{Load sensitivities}}.
\end{aligned}
\end{equation}
The load sensitivities terms for the compliance and the multi-criteria objectives are indicated in Eq.~\eqref{sec3:eq8} and Eq.~\eqref{sec3:eq10}, respectively.
We use a density filter \citep{bruns2001topology,bourdin2001filters} with consistent sensitivities to control the minimum length scale of structural features in the topologically optimized pressure loaded structures and compliant mechanisms. 

\subsection{Verification of the Formulation}\label{sec3.5}
To demonstrate that evaluation of the consistent nodal loads (Sec.~\ref{sec2.2.2}) from the obtained pressure field (Sec.~\ref{sec2.2.1}) produces physically correct results, a test problem for pressure loaded structures (Sec.~\ref{sec3.1}) is considered. 
 \begin{figure}[h!]
 	\centering
 	\includegraphics[scale = 1]{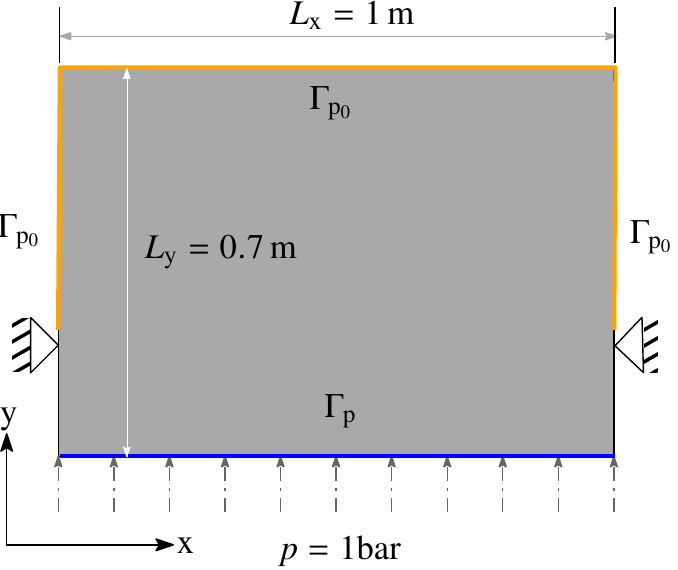}
 	\caption{A design domain for verifying the presented formulation}
 	\label{fig:sec3fig1}
 \end{figure}

	Consider a design domain with dimensions $L_\mathrm{x}=1\si{\meter}$ and $L_\mathrm{y}=0.70\si{\meter}$ in horizontal and vertical directions, respectively (Fig.~\ref{fig:sec3fig1}). The  domain is fixed at locations  $\bm{x} =(0,\,0.3)\si{\meter}$ and $\bm{x} =(1,\,0.3)\si{\meter}$. To discretize the domain, $N_\mathrm{ex}=10$ and $N_\mathrm{ey}=7$ quadrilateral bilinear FEs are used in horizontal and vertical directions respectively. This low resolution mesh is used here to better illustrate the resulting pressure field and nodal forces, more representative numerical examples with finer meshes follow in the next section. A prescribed pressure $p$ of $1\si{\bar}$ i.e. $\SI{1e5}{\newton\meter\tothe{-2}}$ is applied to the bottom (Fig.~\ref{fig:sec3fig1}). The out-of-plane thickness is set to $t=\SI{0.01}{\meter}$ and a plane-stress condition is used. Evidently (Fig.~\ref{fig:sec3fig1}), prior to analysis, the force contribution from the prescribed pressure appears only in y$-$direction with magnitude $p\times t\times L_\mathrm{x} = \SI{1000}{\newton}$.
	
\begin{figure}[h!]
		\begin{subfigure}[t]{0.45\textwidth}
			\centering
			\includegraphics[scale=0.9]{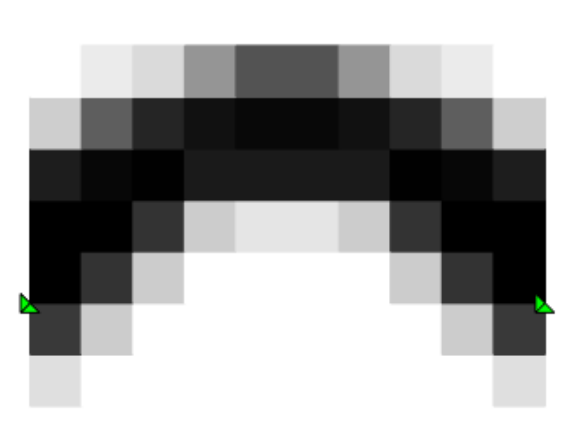}
			\caption{}
			\label{fig:sec3fig2a}
		\end{subfigure}
	\quad
		\begin{subfigure}[t]{0.45\textwidth}
			\centering
			\includegraphics[scale=1]{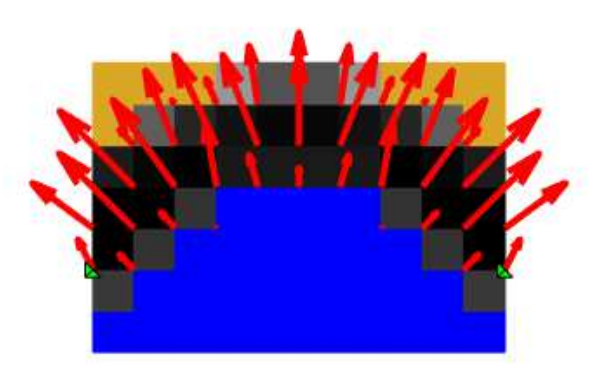}
			\caption{}
			\label{fig:sec3fig2b}
		\end{subfigure}
		\caption{(\subref{fig:sec3fig2a}) The final continuum (\subref{fig:sec3fig2b}) The final continuum with pressure field and nodal force distribution. The obtained resultant forces in $x-$ and $y-$directions are $\SI{0}{\newton}$ and $\SI{1000}{\newton}$, respectively. The resultant force at initial and final state has same direction ($+y$) and magnitude ($\SI{1000}{\newton}$). The developed pressure field inside the given domain is indicated in blue, and regions with pressure $p_\mathrm{out}$ are indicated by orange.}\label{fig:sec3fig2} 
	\end{figure}

	A  linear material model with Young's modulus $E = \SI{3e9}{\newton\per\square\meter}$ and Poisson's ratio $\nu =0.4$ is considered. The other optimization parameters such as penalization parameter $\zeta$,  minimum Young's modulus $E_\mathrm{min}$ and the Darcy parameters are listed in Table \ref{Table:T1} (Sec.~\ref{Sec4}). The filter radius and volume fraction are set to $1.2\times\min({\frac{L_\mathrm{x}}{N_\mathrm{ex}},\,\frac{L_\mathrm{y}}{N_\mathrm{ey}}})$ and $0.45$, respectively. The volume fraction is used to initialize all density variables. Furthermore, the parameter $h_\mathrm{s}$ is evaluated using Eq.~(\ref{sec2:eq5}) with $r =0.1$ and $\Delta s = 2\times\max(\frac{L_\mathrm{x}}{N_\mathrm{ex}},\,\frac{L_\mathrm{y}}{N_\mathrm{ey}})$. The MMA optimizer \citep{svanberg1987method} is used herein with default settings, except the move limit i.e. change in density is set to $0.1$ in each optimization iteration. The results in Fig.~\ref{fig:sec3fig2} are depicted after 100 MMA optimizer iterations.  
	
	\begin{figure*}[h!]
		\begin{subfigure}[t]{0.225\textwidth}
			\centering
			\includegraphics[scale=0.6]{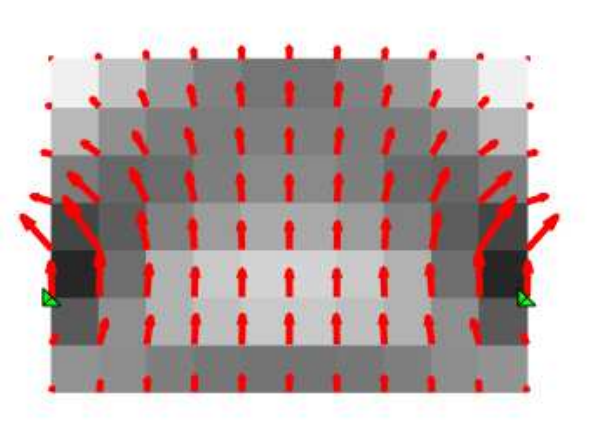}
			\caption{Iteration $5: \mathrm{F_x^r} =\SI{0.0}{\newton}$,\,$\mathrm{F_y^r} =\SI{1000.0}{\newton}$}
			\label{fig:sec3fig3a}
		\end{subfigure}
		\begin{subfigure}[t]{0.225\textwidth}
			\centering
			\includegraphics[scale=0.6]{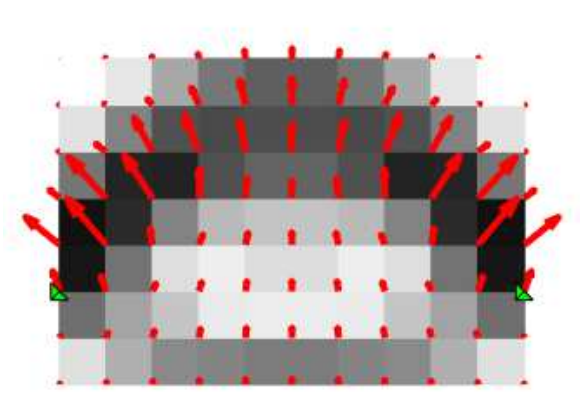}
			\caption{Iteration $10: \mathrm{F_x^r} =\SI{0.0}{\newton}$,\,$\mathrm{F_y^r} =\SI{1000.0}{\newton}$}
			\label{fig:sec3fig3b}
		\end{subfigure}
	\begin{subfigure}[t]{0.225\textwidth}
		\centering
		\includegraphics[scale=0.6]{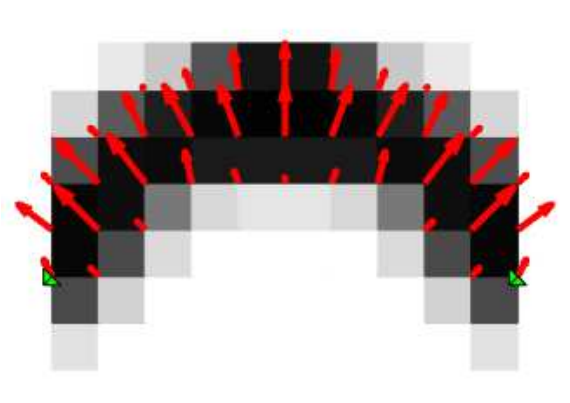}
		\caption{Iteration $15: \mathrm{F_x^r} =\SI{0.0}{\newton}$,\,$\mathrm{F_y^r} =\SI{1000.0}{\newton}$}
		\label{fig:sec3fig3c}
	\end{subfigure}
\begin{subfigure}[t]{0.225\textwidth}
	\centering
	\includegraphics[scale=0.6]{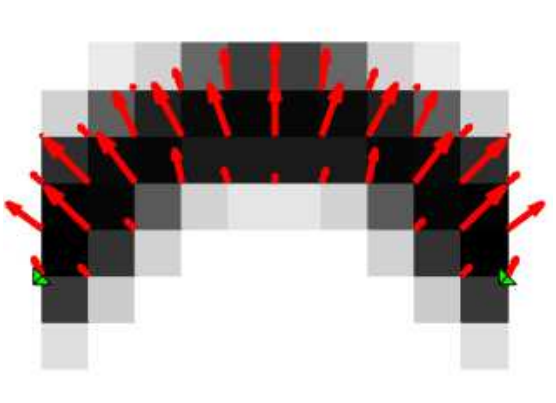}
	\caption{Iteration $20: \mathrm{F_x^r} =\SI{0.0}{\newton}$,\,$\mathrm{F_y^r} =\SI{1000.0}{\newton}$}
	\label{fig:sec3fig3d}
\end{subfigure}
		\caption{Nodal force distribution at different instances of the TO process (iterations). It is found that the resultant force at each instance is same to that of the initial state. \textbf{Key:} $\mathrm{F_x^r}-$ the resultant force in  x$-$direction and $\mathrm{F_y^r}-$ the resultant force in y$-$direction.}\label{fig:sec3fig3}
	\end{figure*}

Fig.~\ref{fig:sec3fig2} depicts the final continuum, pressure field and its nodal force distribution originating from the prescribed pressure at the final state. The pressurized regions are indicated in blue and the low pressure regions are represented by orange. Note that the used color scheme (Fig.~\ref{fig:sec3fig2b}) has been considered for all other numerical problems solved in Sec. \ref{Sec4}. It is found that the magnitude and direction of the resultant force at final and initial state are the same. In addition, they are same in all other instances of the optimization (Fig.~\ref{fig:sec3fig3}). This confirms that the pressure field is correctly converted into consistent nodal loads using the global conversion  matrix $\mathbf{H}$ (Sec. \ref{sec2.2.2}). One can also notice (Fig.~\ref{fig:sec3fig3}), the present method results in spreading of the nodal forces instead of confining them to a narrow (imposed) boundary  as considered in Ref. \citep{Hammer2000,Du2004a,Lee2012a}. This may help the TO process to explore a larger part of the design space and to find a better solution. As the design converges to a 0/1 solution, the region over which the pressure spreads reduces, and thus the loading approaches a boundary load.
 
	\section{Numerical Results and Discussion}\label{Sec4}
	\begin{table*}[h!] 
	\centering
	\begin{tabular}{ l c c }
		\hline
		\hline
		\textbf{Nomenclature} & \textbf{Notation} & \textbf{Value}  \\ \hline
		\textit{Material parameters} &  &  \rule{0pt}{3ex}\\ 
		Young's Modulus & $E$ & $\SI{3e9}{\newton\per\square\meter}$ \rule{0pt}{3ex}\\ 
		Poisson's ratio & $\nu$ & $0.40$\\ \hline
		\textit{Optimization parameters} &  & \\
		Penalization (Eq.~\ref{sec3:eq1}) &$\zeta$ &$3$ \rule{0pt}{3ex}   \\
		Minimum E & $E_\mathrm{min}$ &$E \times 10^{-5} \si{\newton\per\square\meter}$ \\
		Move limit & $\Delta \bm{\rho}$ & 0.1 per iteration\\ \hline
		\textit{Objective parameters} & & \\ 
		Input pressure load &$p_\mathrm{in}$ & $\SI{1e5}{\newton\per\square\meter}$ \rule{0pt}{3ex}\\
		Output spring stiffness & $k_\mathrm{ss}$ &$\SI{1e4}{\newton\per\meter}$\\ \hline
		\textit{Darcy parameters} & & \\ 
		$K(\bm{\rho})$ step location	& 	$\eta_k$ 	& 0.4 \rule{0pt}{3ex}\\
		$K(\bm{\rho})$ slope	at step	& 	$\beta_k$	& 10\\
		$H(\bm{\rho})$ step location	& 	$\eta_h$ 	& 0.6\\
		$H(\bm{\rho})$ slope	at step	& 	$\beta_h$	& 10\\ 
		Conductivity in solid &   $k_\mathrm{s} $  & $\SI{1e-10}{\meter\tothe{4}\per\newton\per\second}$\\
		Conductivity in void	&   $k_\mathrm{v}$  & $\SI{1e-3}{\meter\tothe{4}\per\newton\per\second}$\\
		Drainage from solid	&   $h_\mathrm{s} $  & $\left(\frac{\ln{r}}{\Delta s}\right)^2 k_\mathrm{s}$ \\
		Remainder of input pressure at $\Delta s$ &r& 0.1\\
		Depth wherein the limit $r$ reached &$\Delta s$ & 0.002m\\\hline\hline
	\end{tabular}
	\caption{Various parameters used in the TO examples.}\label{Table:T1}
\end{table*}

\begin{figure*}[h!]
	\begin{subfigure}[t]{0.5\textwidth}
		\centering
		\includegraphics[scale=1]{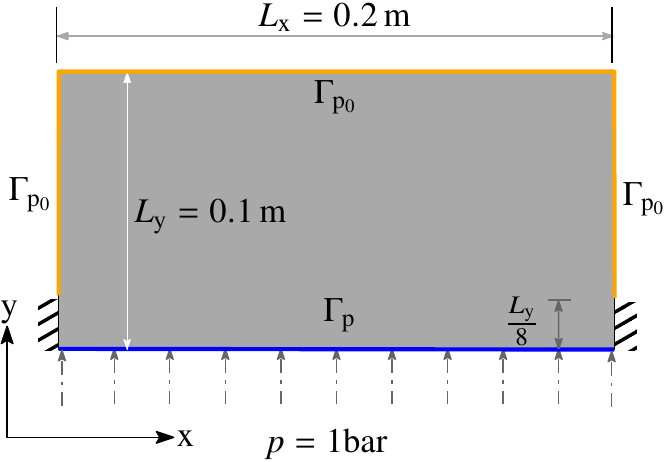}
		\caption{The design domain}
		\label{fig:sec4fig1a}
	\end{subfigure}
	\begin{subfigure}[t]{0.5\textwidth}
		\centering
		\includegraphics[scale=1]{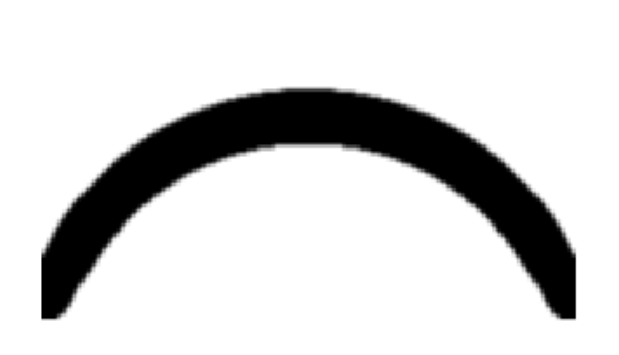}
		\caption{}
		\label{fig:sec4fig1b}
	\end{subfigure}
	\begin{subfigure}[t]{0.5\textwidth}
		\centering
		\includegraphics[scale=1]{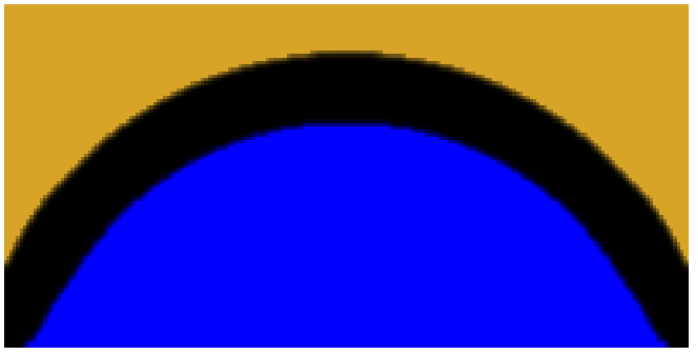}
		\caption{}
		\label{fig:sec4fig1c}
	\end{subfigure}
	\begin{subfigure}[t]{0.5\textwidth}
		\centering
		\includegraphics[scale=0.30]{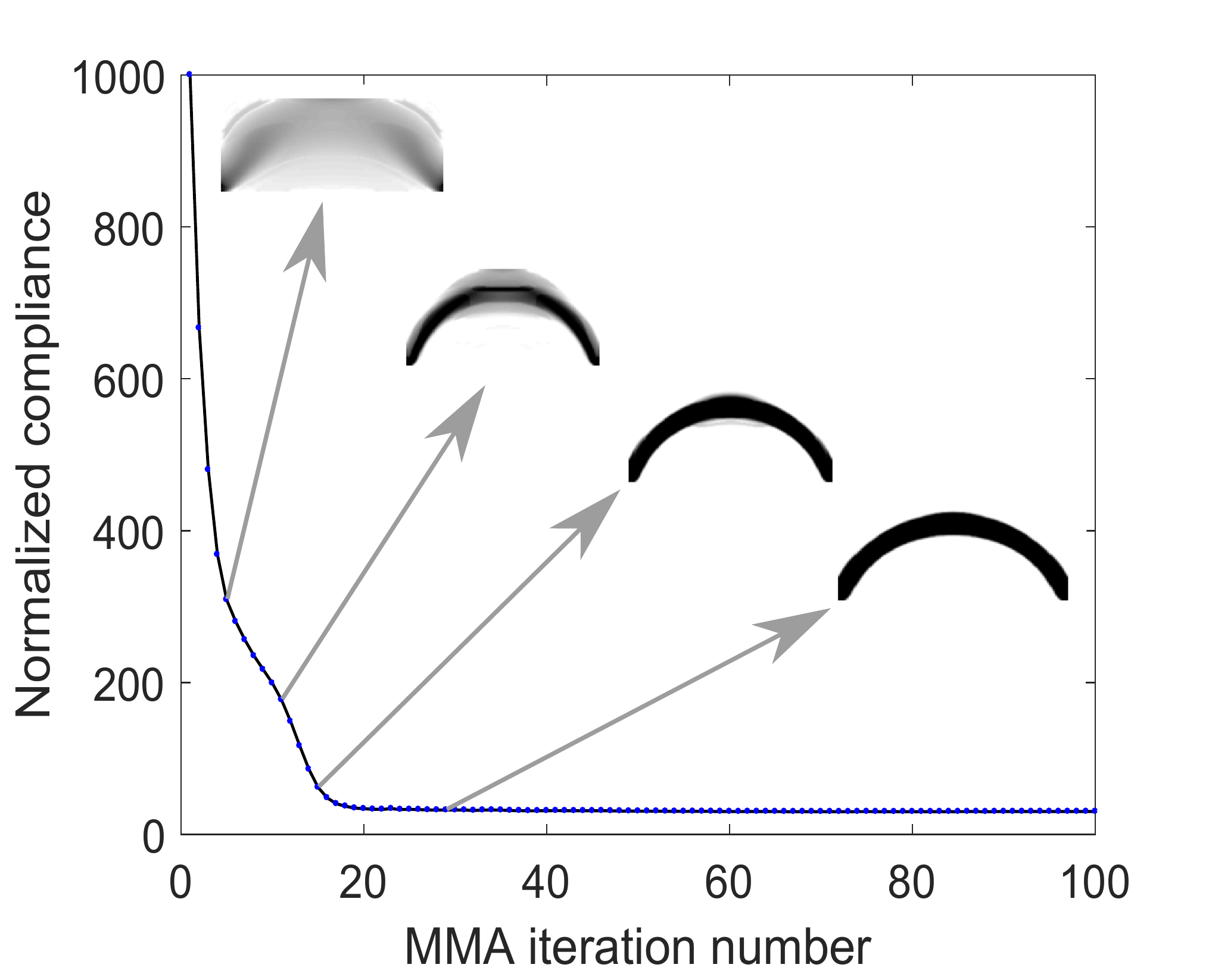}
		\caption{}       
		\label{fig:sec4fig1d}
	\end{subfigure}
	\caption{(\subref{fig:sec4fig1a}) Design domain of size $L_\mathrm{x}\times L_\mathrm{y} = \SI{0.2}{\meter}\times\SI{0.1}{\meter}$ for the internally pressurized arch-structure. A pressure load $p = \SI{1}{\bar}$ is applied on boundary $\mathrm{\Gamma_{p}}$. The fixed displacement boundary and zero pressure boundary $\mathrm{\Gamma_{P_0}}$ are also depicted. Results of the problem, (\subref{fig:sec4fig1b}) Optimized solution, $f_0^\mathrm{s} = \SI{30.27}{\newton\meter}$  (\subref{fig:sec4fig1c}) Optimized solution with pressure field and (\subref{fig:sec4fig1d}) Convergence history with intermediate designs.}\label{fig:sec4fig1}
\end{figure*}
In this section, various (benchmark) design problems involving pressure loaded stiff structures and small deformation compliant mechanisms are solved to show the efficacy and robustness of the present method. Table \ref{Table:T1} depicts the nomenclature, notations and numerical values for different parameters used in the TO. Any change in the value of considered parameters is reported within the definition of the problem formulation. In all the examples presented herein, one design variable per FE is used and topology optimization is initialized using the given volume fraction.
\subsection{Internally pressurized arch-structure}\label{IPAS}

\begin{figure*}[h!]
	\begin{subfigure}[t]{0.5\textwidth}
		\centering
		\includegraphics[scale=1]{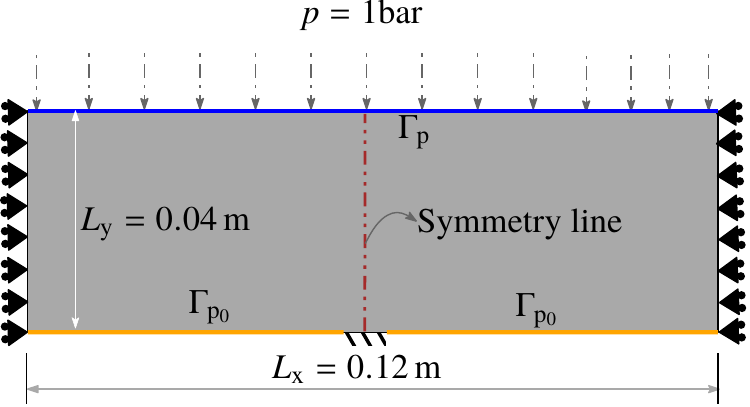}
		\caption{The design domain}
		\label{fig:sec4fig3a}
	\end{subfigure}
	\begin{subfigure}[t]{0.5\textwidth}
		\centering
		\includegraphics[scale=1]{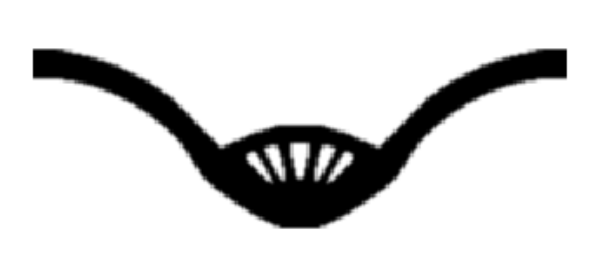}
		\caption{}
		\label{fig:sec4fig3b}
	\end{subfigure}
	\begin{subfigure}[t]{0.5\textwidth}
		\centering
		\includegraphics[scale=1]{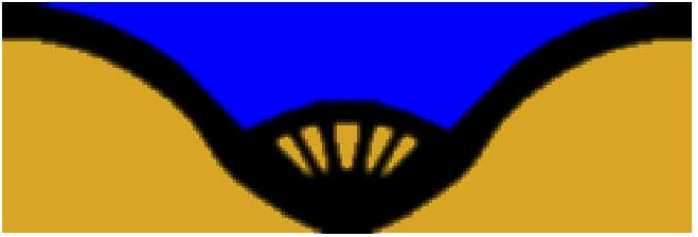}
		\caption{}
		\label{fig:sec4fig3c}
	\end{subfigure}
	\begin{subfigure}[t]{0.5\textwidth}
		\centering
		\includegraphics[scale=0.30]{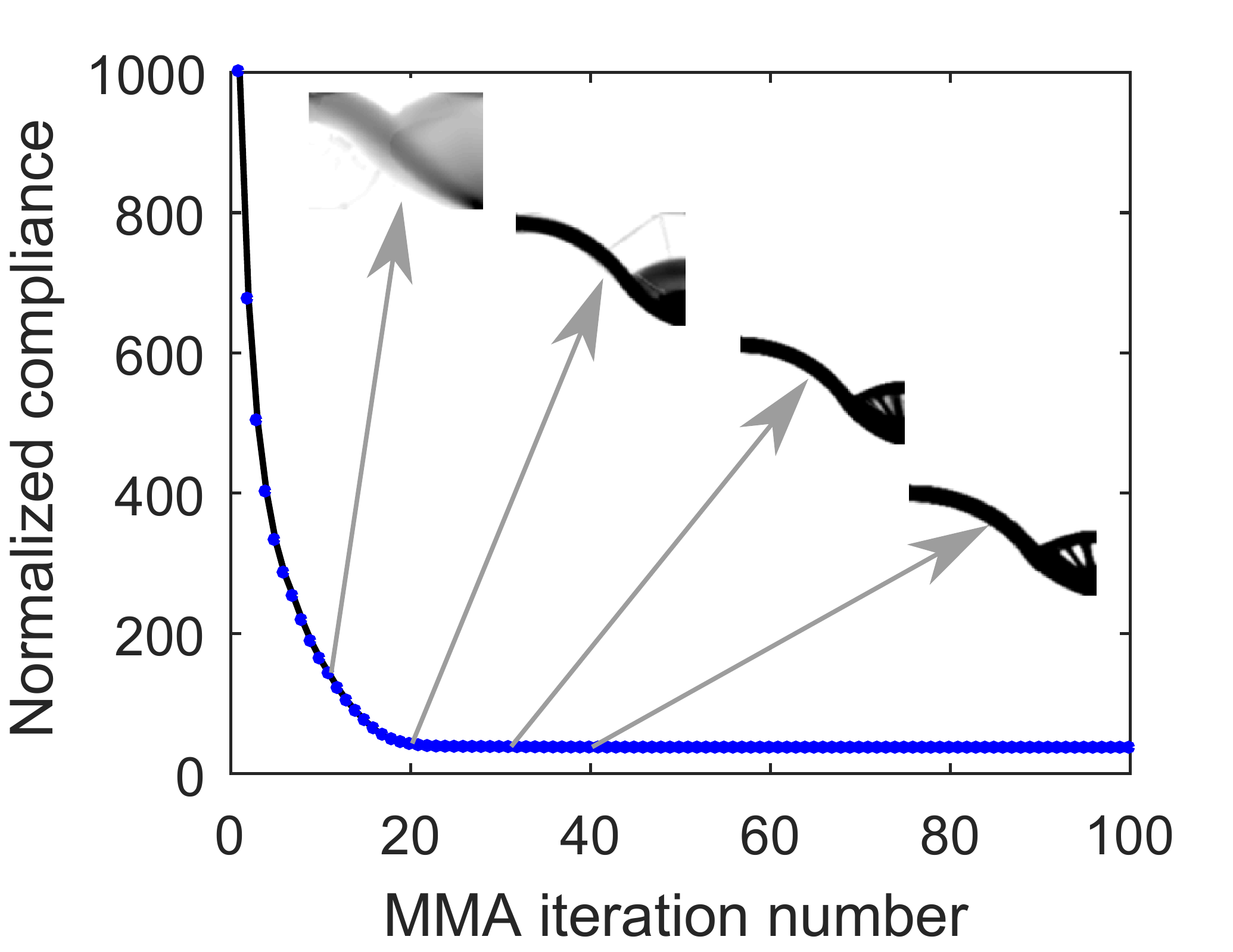}
		\caption{}
		\label{fig:sec4fig3d}
	\end{subfigure}
	\caption{(\subref{fig:sec4fig3a}) Design domain for piston design with pressure load $p = \SI{1}{\bar}$ on boundary $\mathrm{\Gamma_{p}}$, fixed displacement boundary and zero pressure boundary $\mathrm{\Gamma_{P_0}}$. (\subref{fig:sec4fig3b}) Optimized solution, $f_0^\mathrm{s} = \SI{35.39}{\newton\meter}$ (\subref{fig:sec4fig3c}) Optimized solution with pressure field and nodal forces in red arrows and (\subref{fig:sec4fig3d}) Convergence history with symmetrically half intermediate designs.}\label{fig:sec4fig3}
\end{figure*}
\begin{figure*}[h!]
	\begin{subfigure}[t]{0.5\textwidth}
		\centering
		\includegraphics[scale=1]{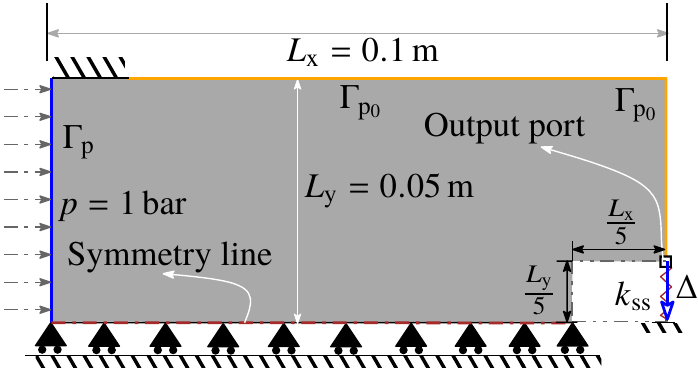}
		\caption{The symmetric half design domain}
		\label{fig:sec4fig5a}
	\end{subfigure}
	\begin{subfigure}[t]{0.5\textwidth}
		\centering
		\includegraphics[scale=0.65]{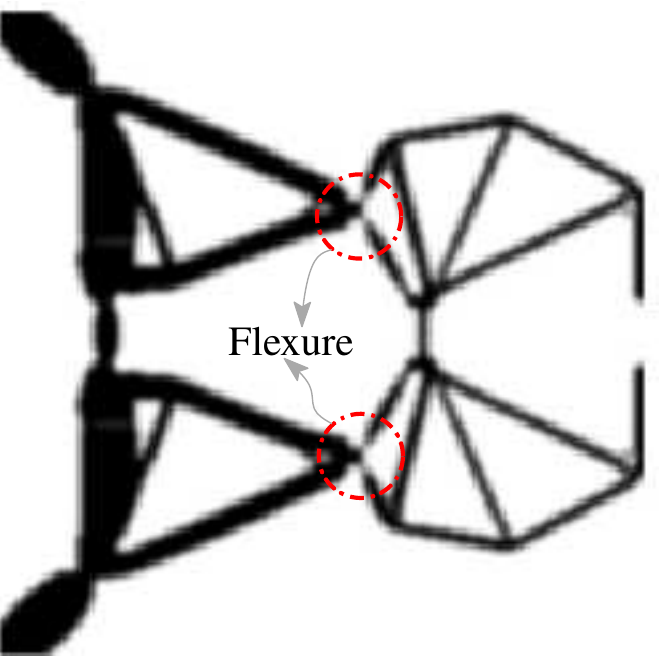}
		\caption{$f_0^\mathrm{CM} = {-1013.6},\,\Delta = \SI{0.287}{\milli\meter}$}
		\label{fig:sec4fig5b}
	\end{subfigure}
	\begin{subfigure}[t]{0.5\textwidth}
		\centering
		\includegraphics[scale=0.65]{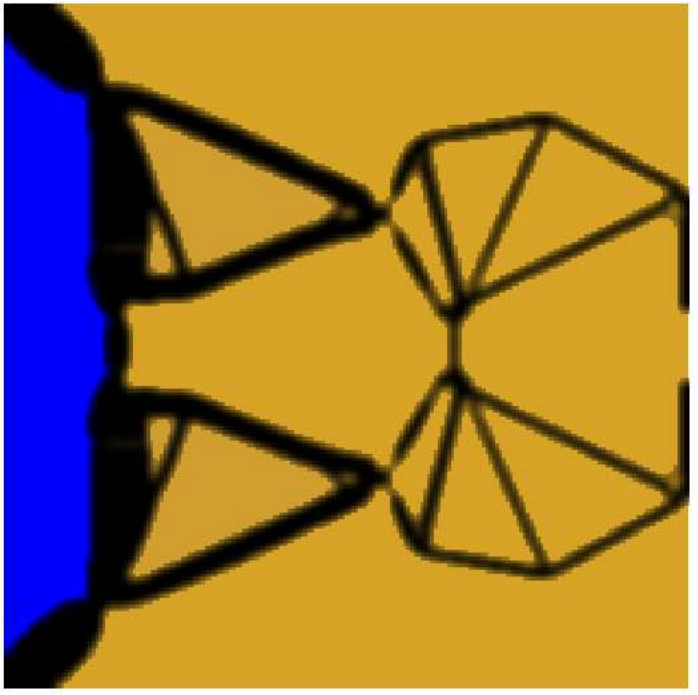}
		\caption{Solution with pressure field}
		\label{fig:sec4fig5c}
	\end{subfigure}
	\begin{subfigure}[t]{0.5\textwidth}
		\centering
		\includegraphics[scale=0.3]{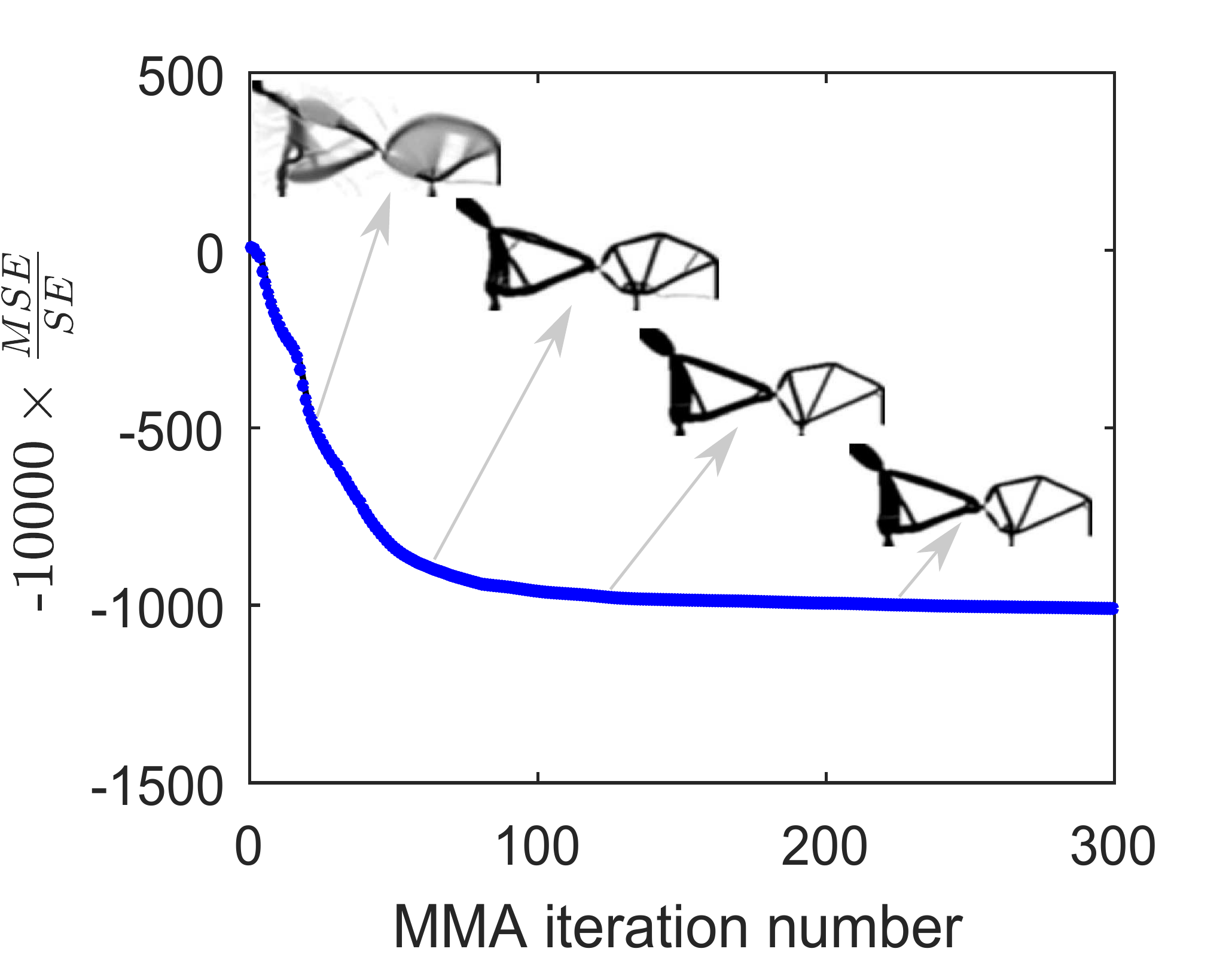}
		\caption{Convergence history}
		\label{fig:sec4fig5d}
	\end{subfigure}
	\caption{(\subref{fig:sec4fig5a}) Half design domain for crimper mechanism. The figure shows the pressure loading boundary $\mathrm{\Gamma_{b}}$ with pressure $p = \SI{1}{\bar}$, fixed displacement boundary, zero pressure boundary $\mathrm{\Gamma_{P_0}}$, symmetry line, output port and the direction of the desired deformation $\mathrm{\Delta}$.  (\subref{fig:sec4fig5b}) Optimized crimper mechanism (\subref{fig:sec4fig5c}) Optimized crimper mechanism with pressure field and (\subref{fig:sec4fig5d}) Convergence history of the problem with some intermediate designs at different instances of the TO.}\label{fig:sec4fig5}
\end{figure*}
In this example that was introduced in \cite{Hammer2000}, a structure subjected to a pressure load $p=\SI{1}{\bar}$ from the bottom is designed by  minimizing its compliance (Eq.~\ref{sec3:eq2}). The design domain is sketched in Fig.~\ref{fig:sec4fig1a}. The dimensions in x and y directions are  $L_\mathrm{x} = \SI{0.2}{\meter}$ and $L_\mathrm{y}= \SI{0.1}{\meter}$, respectively. The bottom part of left and right sides of the domain is fixed as depicted in Fig.~\ref{fig:sec4fig1a}. $\mathrm{\Gamma_{P_0}}$ indicates boundary with zero pressure. 

$N_\mathrm{ex}\times N_\mathrm{ey} =200\times100$ quad-elements are employed to discretize the domain, where $N_\mathrm{ex}$ and $N_\mathrm{ey}$ are number of quad-FEs in horizontal and vertical directions, respectively. Out-of-plane thickness is set to $t=\SI{0.01}{\meter}$ with plane-stress condition.  The volume fraction is set to $0.25$. The filter radius is set to $2\times\min({\frac{L_\mathrm{x}}{N_\mathrm{ex}},\,\frac{L_\mathrm{y}}{N_\mathrm{ey}}})$. The Young's modulus and Poission's ratio are set to $\SI{3e9}{\newton\per\square\meter}$ and $0.40$ respectively. Other parameters such as material parameters, optimization parameters and Darcy parameters are same as mentioned in Table~\ref{Table:T1}. 

The final continuum after $100$ MMA optimization iterations is depicted in Fig.~\ref{fig:sec4fig1b}, with the normalized objective $f_0^\mathrm{s} = \SI{30.27}{\newton\meter}$. The topology of the result is similar to that obtained in previous literature, e.g., Refs. \citep{Hammer2000,Du2004a}. The final continuum with pressure field is shown in Fig.~\ref{fig:sec4fig1c}. The color scheme for the pressure field is as mentioned in Sec. \ref{Sec3}. The convergence history plot with evolving designs at some instances of the TO is depicted in Fig.~\ref{fig:sec4fig1d}. Smooth and relatively rapid convergence is observed. It is noted that from a relatively diffused initial interface, the boundary exposed to pressure loading is gradually formed during the optimization process.

\subsection{Piston}
 The design with dimension $L_\mathrm{x}\times L_\mathrm{y} = \SI{0.12}{\meter}\times\SI{0.04}{\meter}$ of a piston for a general mechanical application is shown in Fig.~\ref{fig:sec4fig3a}. The figure  depicts the design specification, pressure boundary loading, fixed boundary/location and a vertical symmetry line. It is desired to find a stiffest optimum continuum which can convey the applied pressure loads on the upper boundary to the lower fixed support readily (Fig.~\ref{fig:sec4fig3a}). We exploit the symmetry present in the domain to find the optimum solution. The problem was originally introduced and solved in \cite{bourdin2003design}.

The symmetric half  of the domain is parameterized using $N_\mathrm{ex}\times N_\mathrm{ey} = 120\times80$ number of the standard quad-elements. Volume fraction is set to $V^* = 0.25$. The density filter radius is  $1.8\times\min({\frac{L_\mathrm{x}}{N_\mathrm{ex}},\,\frac{L_\mathrm{y}}{N_\mathrm{ey}}})$. The Young's modulus, Poission's ratio, and out-of-plane thickness are kept same as those of arch-structure design. $\eta_\mathrm{k},\,\beta_\mathrm{k},\,\eta_\mathrm{h}$ and $\beta_\mathrm{h}$ are set to $0.20,\,10,\,0.30$ and $10$, respectively. Other required design variables are same as mentioned in Table \ref{Table:T1}. 

Fig.~\ref{fig:sec4fig3b} depicts the optimum solution to the problem after 100 iterations of the MMA optimizer. The normalized compliance of the structure at this stage is equal to $f_0^\mathrm{s}= \SI{35.39}{\newton\meter}$. The obtained topology closely resembles those found in Refs. \citep{Lee2012a,Wang2016,Picelli2019} for similar problems with different design and optimization settings. The optimized continuum with pressure field is shown in Fig.~\ref{fig:sec4fig3c}. The convergence history plot for symmetric half design is depicted in Fig.~\ref{fig:sec4fig3d}. 

\begin{figure*}[h!]
	\begin{subfigure}[t]{0.5\textwidth}
		\centering
		\includegraphics[scale=1]{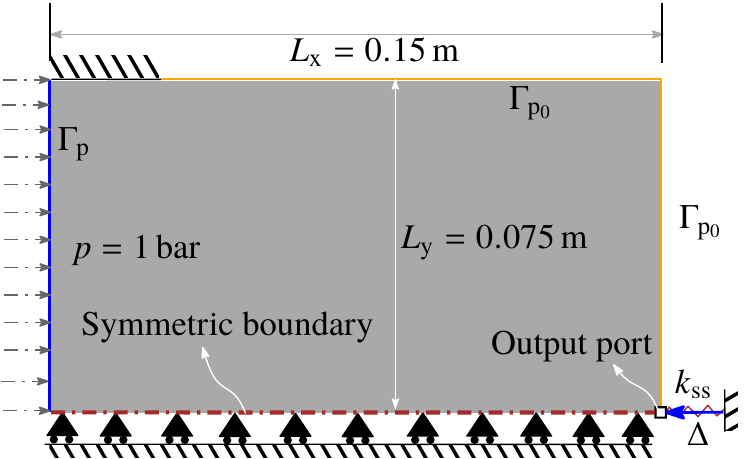}
		\caption{The symmetric half design domain}
		\label{fig:sec4fig7a}
	\end{subfigure}
	\begin{subfigure}[t]{0.5\textwidth}
		\centering
		\includegraphics[scale=0.65]{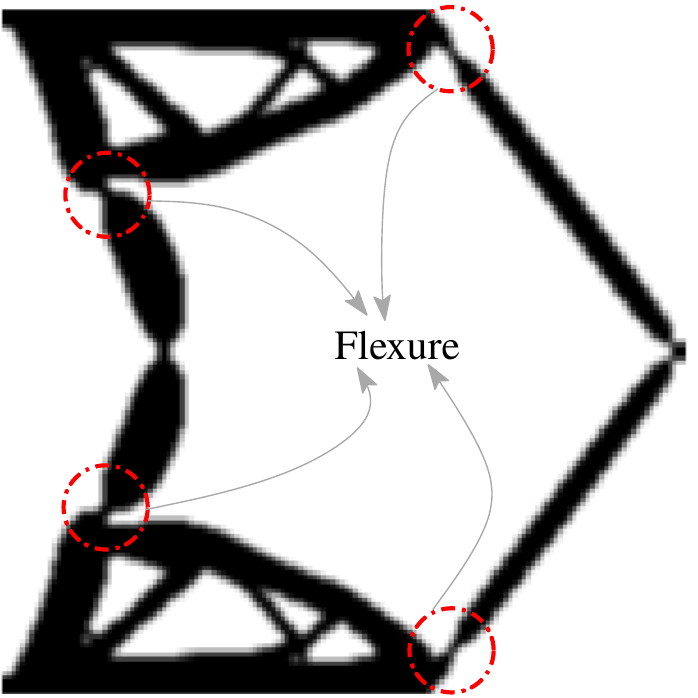}
		\caption{$f_0^\mathrm{CM} = {-369.65},\,\mathrm{\Delta} = \SI{0.221}{\milli\meter}$}
		\label{fig:sec4fig7b}
	\end{subfigure}
	\begin{subfigure}[t]{0.5\textwidth}
		\centering
		\includegraphics[scale=0.75]{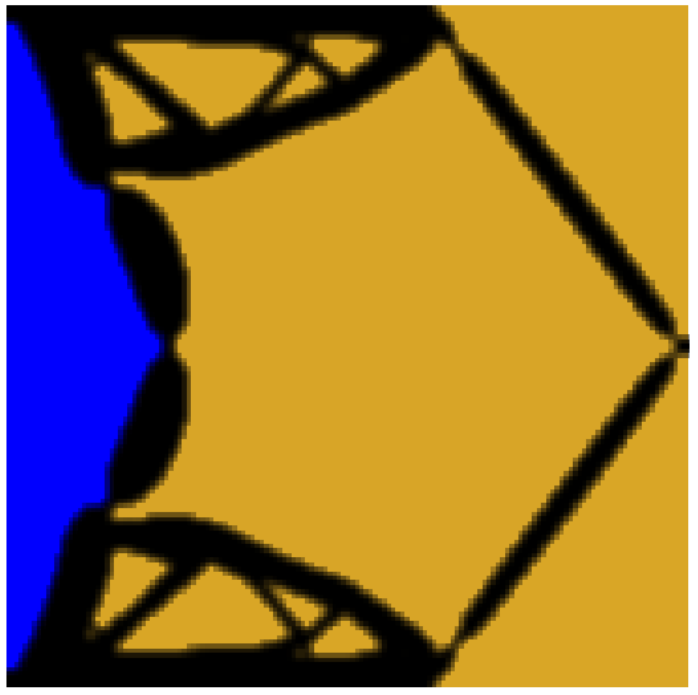}
		\caption{Solution with the pressure field}
		\label{fig:sec4fig7c}
	\end{subfigure}
	\begin{subfigure}[t]{0.5\textwidth}
		\centering
		\includegraphics[scale=0.30]{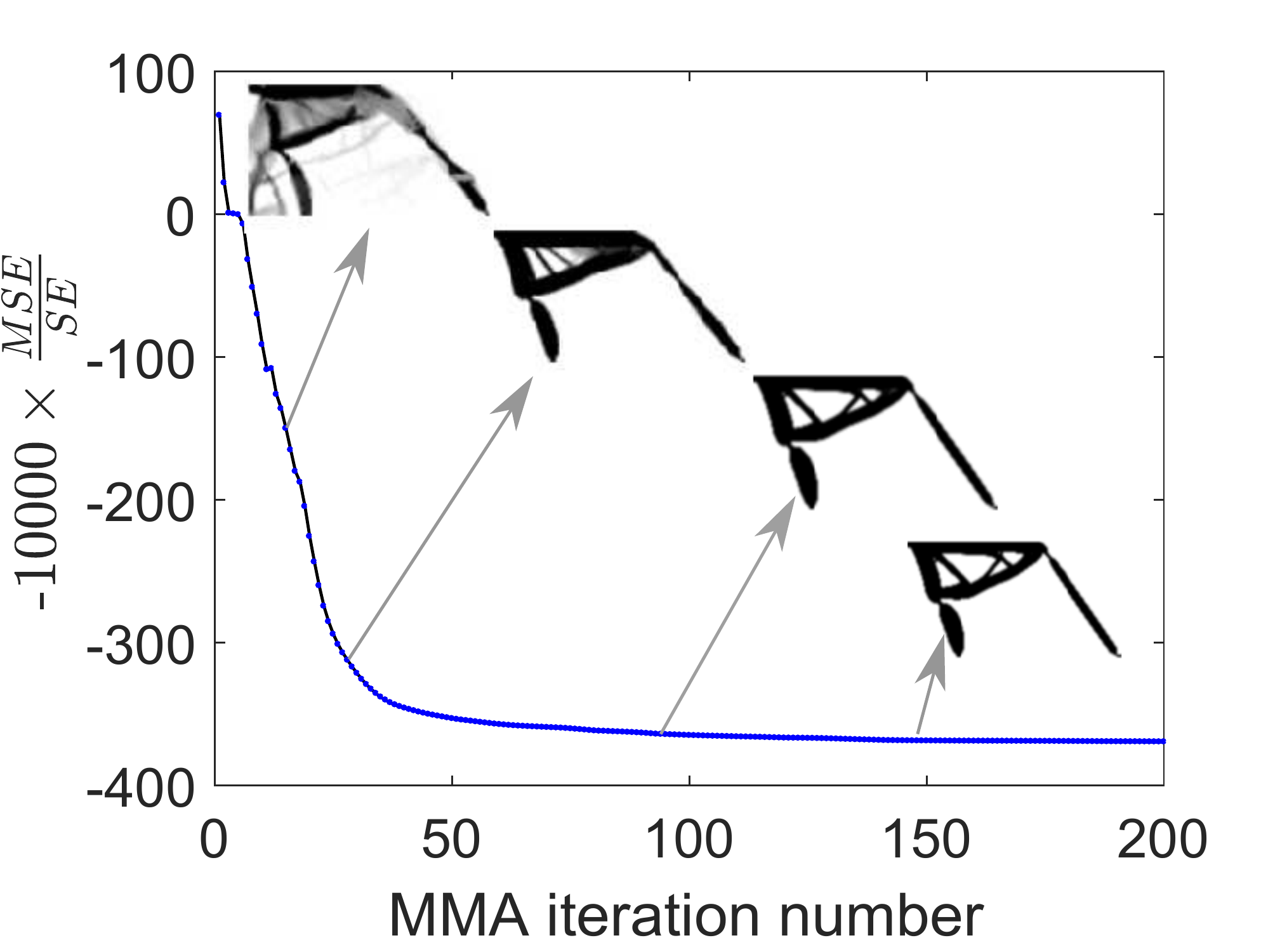}
		\caption{Convergence history plot}
		\label{fig:sec4fig7d}
	\end{subfigure}
	\caption{(\subref{fig:sec4fig7a}) Half design domain for inverter mechanism. The figure depicts the pressure loading boundary $\mathrm{\Gamma_{b}}$ with pressure $p = \SI{1}{\bar}$, fixed displacement boundary, zero pressure boundary $\mathrm{\Gamma_{P_0}}$, symmetric boundary condition and output point.  (\subref{fig:sec4fig7b}) Optimized inverter mechanism (\subref{fig:sec4fig7c}) Optimized inverter mechanism with pressure field and (\subref{fig:sec4fig7d}) Convergence history plot of the problem with some intermediate designs.}\label{fig:sec4fig7}
\end{figure*}

\begin{figure*}[h!]
	\begin{subfigure}[t]{0.5\textwidth}
		\centering
		\includegraphics[scale=0.75]{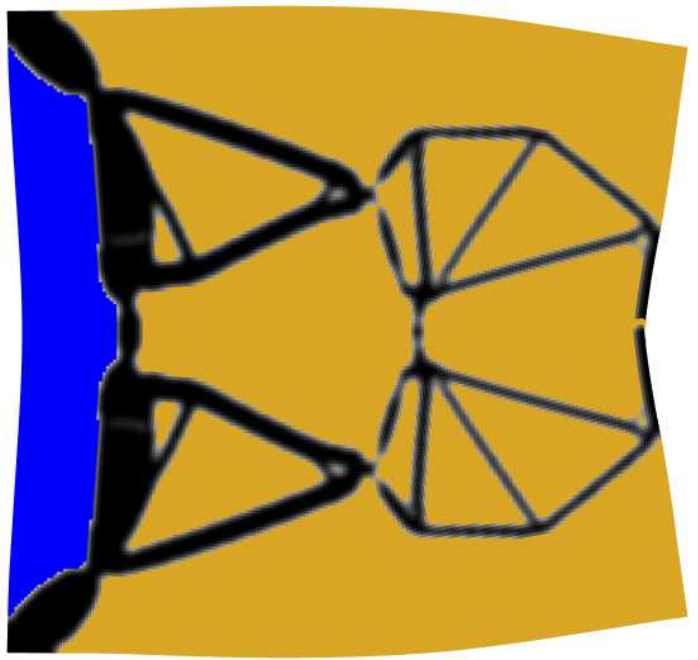}
		\caption{Deformed Compliant Crimper Mechanism}
		\label{fig:sec4fig10a}
	\end{subfigure}
	\begin{subfigure}[t]{0.5\textwidth}
		\centering
		\includegraphics[scale=0.75]{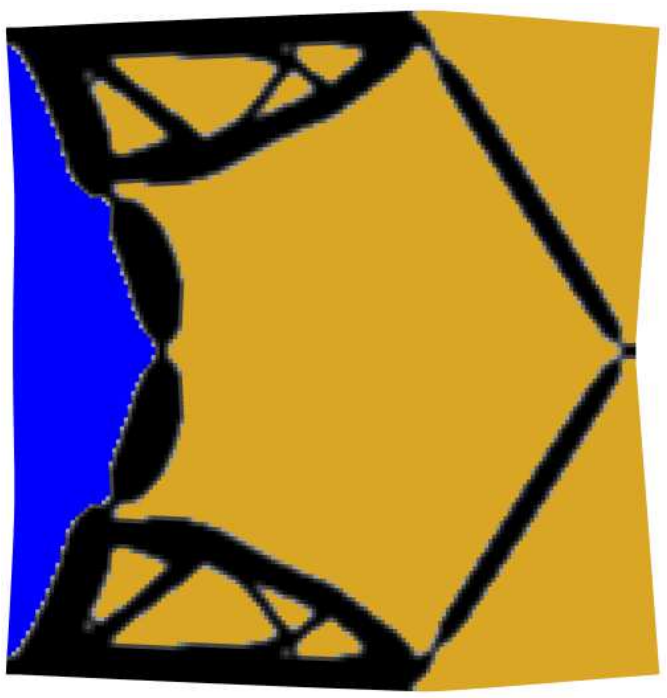}
		\caption{Deformed Compliant Inverter Mechanism}
		\label{fig:sec4fig10b}
	\end{subfigure}
	\caption{The respective actual deformations of CMs are magnified by 20 times to ease visibility of the deformed profiles.}\label{fig:sec4fig10}
\end{figure*}
\begin{figure*}[h!]
	\begin{subfigure}[t]{0.45\textwidth}
		\centering
		\includegraphics[scale=1]{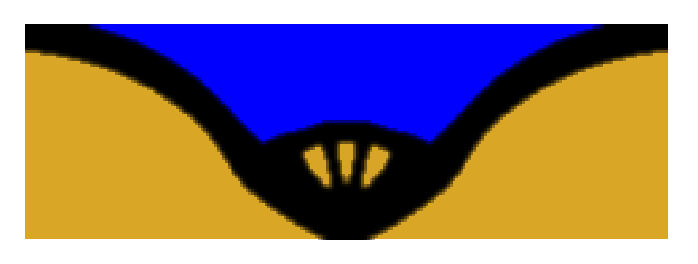}
		\caption{}
		\label{fig:sec4fig11a}
	\end{subfigure}
   \begin{subfigure}[t]{0.5\textwidth}
	\centering
	\includegraphics[scale=0.325]{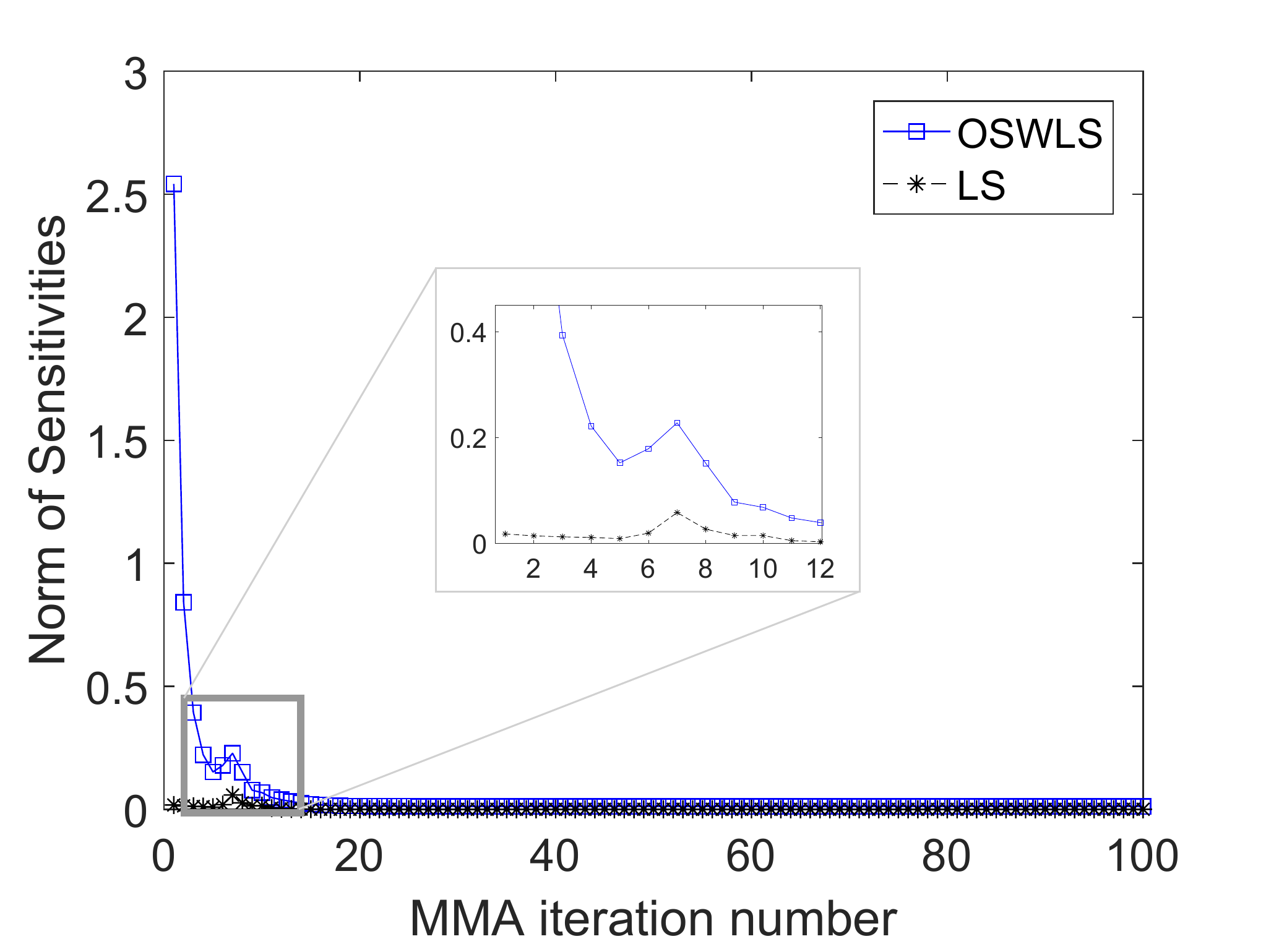}
	\caption{}
	\label{fig:sec4fig11b}
   \end{subfigure}
\begin{subfigure}[t]{0.45\textwidth}
	\centering
	\includegraphics[scale=0.75]{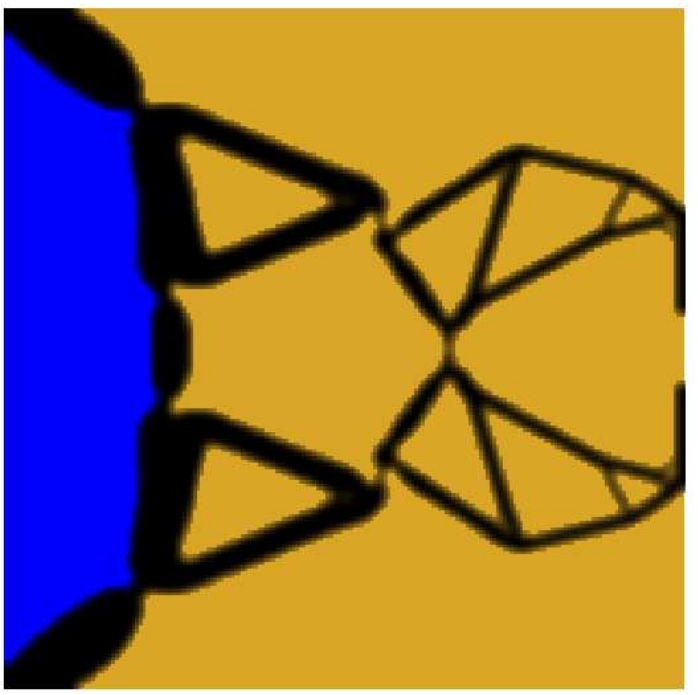}
	\caption{}
	\label{fig:sec4fig11c}
\end{subfigure}
   \begin{subfigure}[t]{0.6\textwidth}
	\centering
	\includegraphics[scale=0.5]{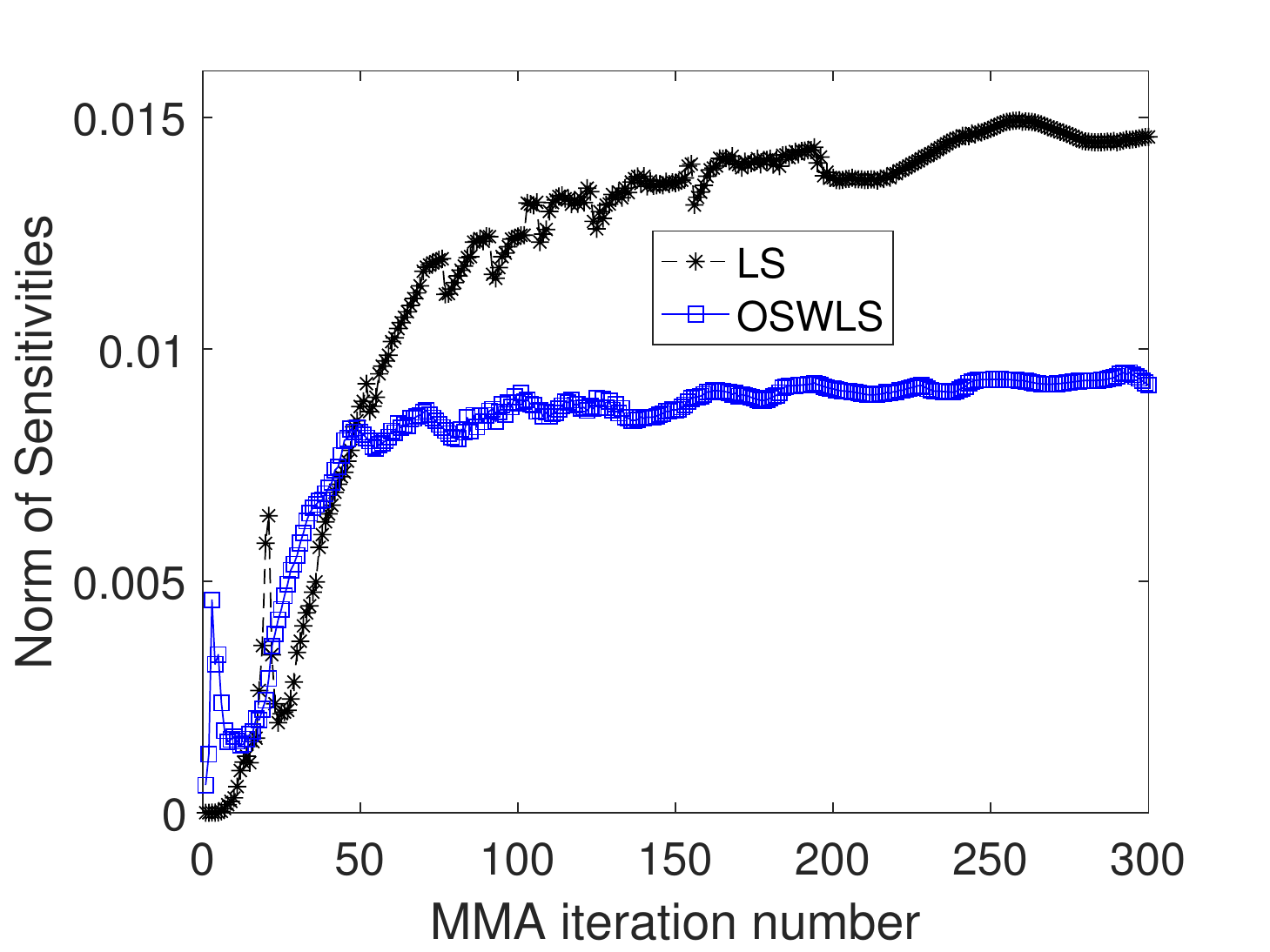}
	\caption{}
	\label{fig:sec4fig11d}
   \end{subfigure}
	\caption{(\subref{fig:sec4fig11a}) Optimized piston design without LS (\subref{fig:sec4fig11b}) Plot of the magnitude ($L_2$-norm) of LS and that of compliance sensitivities without load sensitvities (\subref{fig:sec4fig11c}) Optimized compliant crimper mechanism without LS and (\subref{fig:sec4fig11d}) Plot for magnitude of the LS and that of multi-criteria OSWLS. LS: load sensitivities, OSWLS: objective sensitivities without load sensitivities.\label{fig:sec4fig11}}
\end{figure*}
\subsection{Compliant Crimper Mechanism}
In this  example, a pressure-actuated small deformation compliant crimper is designed. The multi-objective criterion (Eq.~\ref{sec3:eq3}) \citep{saxena2000optimal} is used herein with volume constraint to obtain the optimized compliant crimper. It is desired that pressure acting on the boundary $\mathrm{\Gamma_{p_b}}$ should be transfered to the output port in a manner that the symmetric half of the crimper experiences downward movement at the output port (Fig.~\ref{fig:sec4fig5a}). The design domain for a symmetric half crimper is depicted in Fig.~\ref{fig:sec4fig5a} with associated loading, boundary conditions and other relevant information. Length and width of the depicted domain are $L_\mathrm{x} = \SI{0.1}{\meter}$ and $L_\mathrm{y}=\SI{0.05}{\meter}$, respectively. $t=\SI{0.01}{\meter}$ is taken as the out-of-plane thickness.  Near the output, a void region of area $(\frac{L_\mathrm{x}}{5}\times\frac{L_\mathrm{y}}{5})\si{\meter\squared}$ exists for gripping of a workpiece. However, the domain is parameterized using $N_\mathrm{ex}\times N_\mathrm{ey} = 200\times 100$ bilinear quad-elements considering the domain of size $L_\mathrm{x}\times L_\mathrm{y}$. The FEs present in the void region are set as passive elements with density $\rho= 0$ throughout the simulation. 

Herein, to design the crimper, the volume fraction $V^*$ is taken to $0.20$.  A dummy load of magnitude $\SI{1}{\newton}$ is applied in the direction of the desired deformation at the output port (Fig.~\ref{fig:sec4fig5a}) to evaluate the mutual strain energy (Eq.~\ref{sec3:eq3}). An output  spring of  $k_\mathrm{ss}=\SI{1e4}{\newton\per\meter}$ is attached at the output location, which represents the work-piece stiffness. Filter radius $r_\mathrm{min}= 3\times\min({\frac{L_\mathrm{x}}{N_\mathrm{ex}},\,\frac{L_\mathrm{y}}{N_\mathrm{ey}}})$ is considered. A scaling factor of $10,000$ is used for the objective (Eq.~\ref{sec3:eq3}). Note that the sensitivity of the objective with respect to the design variables is also scaled accordingly. Other design parameters are as mentioned in Table \ref{Table:T1}. 

The symmetric half compliant crimper is solved using the appropriate symmetric condition. We use $300$ MMA iterations. The scaled objective of the mechanism at this stage is $f_0^\mathrm{CM}= -1013.6$ and the recorded output displacement in the required direction is $\Delta = \SI{0.287}{\milli\meter}$. The symmetric half solution is mirrored and combined to get the full solution. Fig.~\ref{fig:sec4fig5b} depicts the solution. The result with pressure field is shown via Fig.~\ref{fig:sec4fig5c}. Fig.~\ref{fig:sec4fig5d} illustrates the convergence history plot with some intermediate designs. Note that the shape of the interface region where pressure is applied to the mechanism evolves during the optimization process. A few gray elements are present in the optimum result, especially near the flexure locations which are relatively thinner (encircled in red, Fig.~\ref{fig:sec4fig5b}) where the deformation is expected to be relatively large.  The TO algorithm prefers flexures at those locations as they allow for large displacement at the output point with marginal strain energy. The robust formulation presented in \cite{wang2011projection} can be used to alleviate such flexures. However, this is not implemented herein, as the motive of the manuscript is to present a novel approach for various pressure-loaded/actuated structure and mechanism problems. The deformed profile of the pressure-actuated compliant crimper mechanism is shown in Fig.~\ref{fig:sec4fig10a}.

\begin{figure*}[h!]
	\begin{subfigure}[t]{0.45\textwidth}
		\centering
		\includegraphics[scale=1]{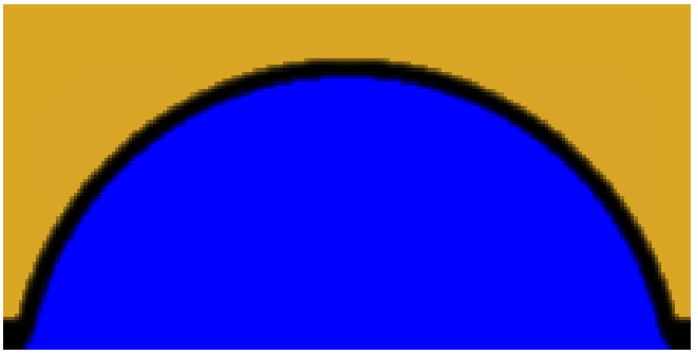}
		\caption{$f_0^\mathrm{s} = \SI{3.31}{\newton\meter},\,V^*=0.075$}
		\label{fig:sec4fig2a}
	\end{subfigure}
	\begin{subfigure}[t]{0.45\textwidth}
		\centering
		\includegraphics[scale=1]{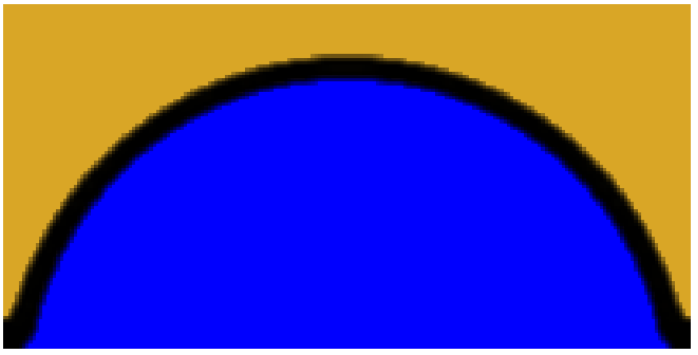}
		\caption{$f_0^\mathrm{s} = \SI{5.46}{\newton\meter},\,V^*=0.1$}
		\label{fig:sec4fig2b}
	\end{subfigure}
	\begin{subfigure}[t]{0.45\textwidth}
		\centering
		\includegraphics[scale=1]{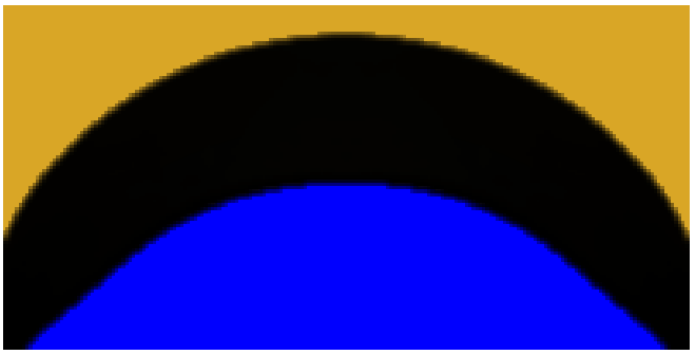}
		\caption{$f_0^\mathrm{s} = \SI{109.51}{\newton\meter},\,V^*=0.45$}
		\label{fig:sec4fig2c}
	\end{subfigure}
	\begin{subfigure}[t]{0.5\textwidth}
		\centering
		\includegraphics[scale=1]{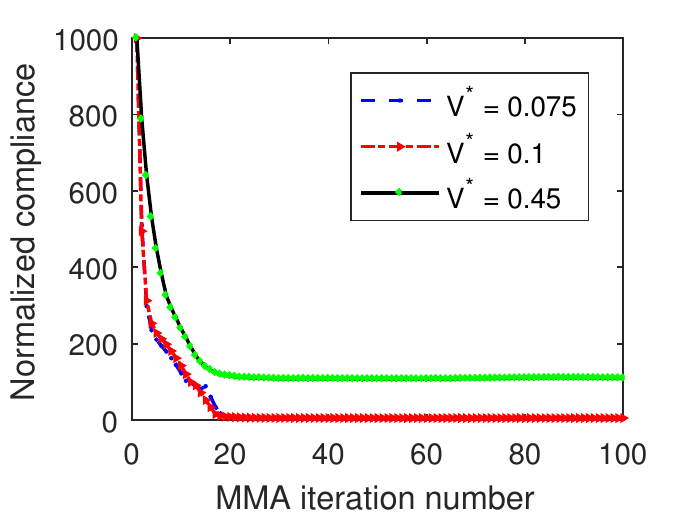}
		\caption{Convergence history}
		\label{fig:sec4fig2d}
	\end{subfigure}
	\caption{Solutions to Example 1 obtained using volume fractions $0.075$ (\subref{fig:sec4fig2a}), $0.01$ (\subref{fig:sec4fig2b}) and $0.45$ (\subref{fig:sec4fig2c}). The optimum continua are shown with respective pressure fields. These solutions are obtained after 100 iterations of the MMA optimizer. (\subref{fig:sec4fig2d}) The convergence history plot for the considered volume fractions.}\label{fig:sec4fig2}
\end{figure*}
\begin{figure*}[h!]
	\begin{subfigure}[t]{0.30\textwidth}
		\centering
		\includegraphics[scale=0.8]{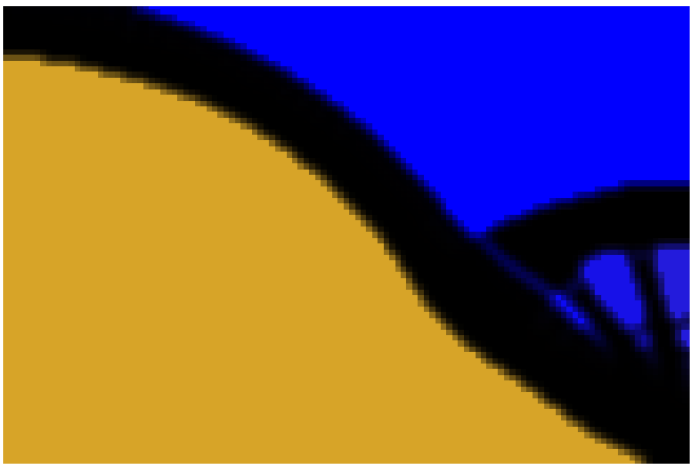}
		\caption{$\beta_k =10,\,\beta_h=10,\,\eta_k =0.4,\,\eta_h = 0.3,\,f_0^\mathrm{s}=\SI{35.13}{N m}$}
		\label{fig:sec4fig6a}
	\end{subfigure}
	\qquad
	\begin{subfigure}[t]{0.30\textwidth}
		\centering
		\includegraphics[scale=0.8]{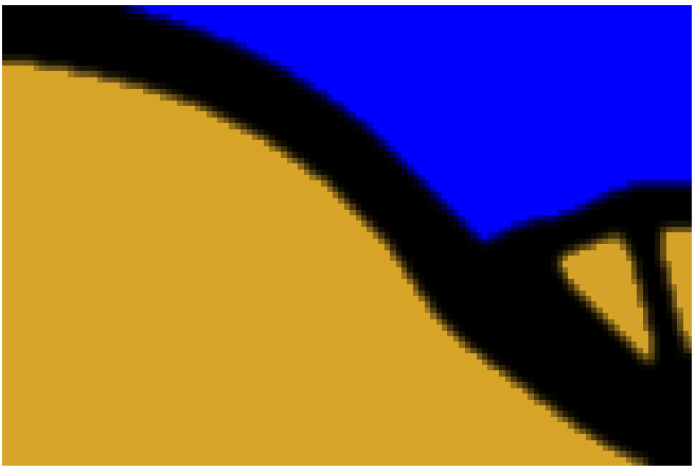}
		\caption{$\beta_k =10,\,\beta_h=10,\,\eta_k =0.4,\,\eta_h = 0.6,\,f_0^\mathrm{s}=\SI{35.03}{N m}$}
		\label{fig:sec4fig6b}
	\end{subfigure}
	\qquad
	\begin{subfigure}[t]{0.30\textwidth}
		\centering
		\includegraphics[scale=0.8]{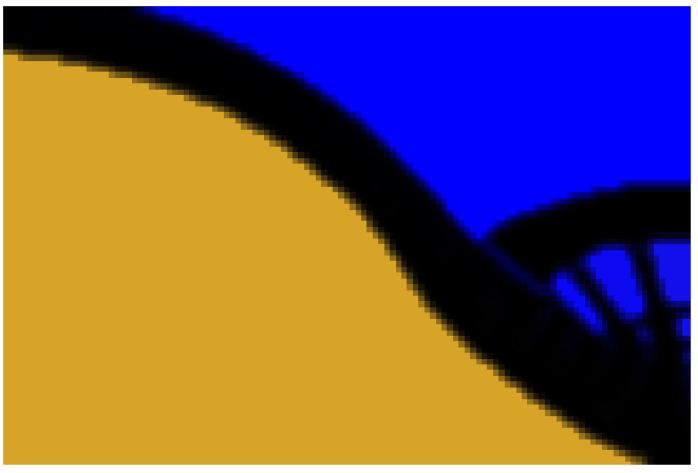}
		\caption{$\beta_k =10,\,\beta_h=15,\,\eta_k =0.4,\,\eta_h = 0.2,\,f_0^\mathrm{s}=\SI{34.79}{Nm}$}
		\label{fig:sec4fig6c}
	\end{subfigure}
	\qquad
	\begin{subfigure}[t]{0.30\textwidth}
		\centering
		\includegraphics[scale=0.8]{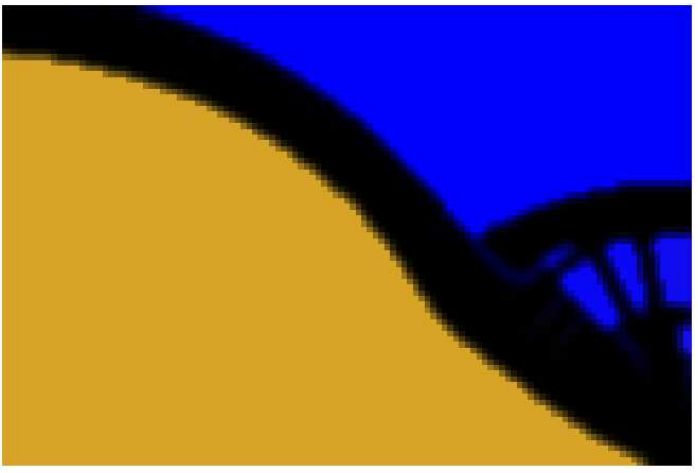}
		\caption{$\beta_k =15,\,\beta_h=15,\,\eta_k =0.6,\,\eta_h = 0.6,\,f_0^\mathrm{s}=\SI{35.04}{Nm}$}
		\label{fig:sec4fig6d}
	\end{subfigure}
	\qquad
	\begin{subfigure}[t]{0.30\textwidth}
		\centering
		\includegraphics[scale=0.8]{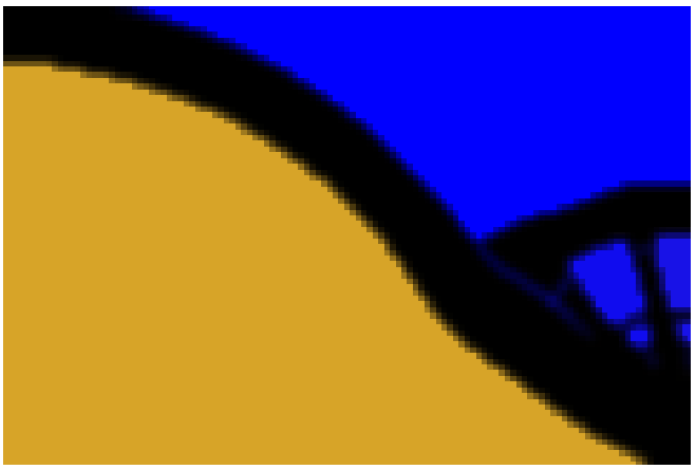}
		\caption{$\beta_k =20,\,\beta_h=20,\,\eta_k =0.6,\,\eta_h = 0.8,\,f_0^\mathrm{s}=\SI{35.11}{Nm}$}
		\label{fig:sec4fig6e}
	\end{subfigure}
	\qquad
	\begin{subfigure}[t]{0.30\textwidth}
		\centering
		\includegraphics[scale=0.8]{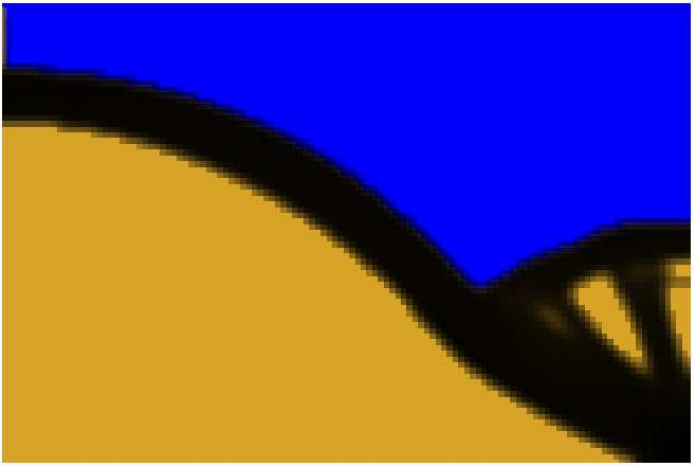}
		\caption{$\beta_k =20,\,\beta_h=20,\,\eta_k =0.2,\,\eta_h = 0.3,\,f_0^\mathrm{s}=\SI{36.91}{Nm}$}
		\label{fig:sec4fig6f}
	\end{subfigure}
	\caption{Solutions to pressure loaded piston design for different conditions}\label{fig:sec4fig6}
\end{figure*}

\subsection{Compliant Inverter Mechanism}\label{CIM}
 A compliant inverter mechanism is synthesized wherein a desired deformation in the opposite direction of the pressure loading  is generated in response to the actuation (Fig.~\ref{fig:sec4fig7a}). The symmetric half design domain with dimensions $L_x = \SI{0.15}{\meter}$ and $L_\mathrm{y}= \SI{0.075}{\meter}$, is depicted in Fig.~\ref{fig:sec4fig7a}. The pressure boundary $\mathrm{\Gamma_{p_b}}$, symmetry boundary, output port and fixed boundary conditions are also indicated via Fig.~\ref{fig:sec4fig7a}. A pressure $p= \SI{1}{\bar}$ is applied on the left side of the design domain. A spring with $k_\mathrm{ss} = \SI{5e4}{\newton\meter}$ representing the reaction force at the output location is taken into account while simulating the problem. The mutual strain energy (Eq.~\ref{sec3:eq3}) is calculated by applying a dummy unit load in the direction of the desired output deformation.

To parametrize the symmetric half design domain, $N_\mathrm{ex}\times N_\mathrm{ey} = 150\times 75$ bilinear quad-elements are employed. The volume fraction $V^*$ is set to $0.25$. The step locations for the flow  $K(\rho)$ and drainage $H(\rho)$ coefficients are set to $\eta_k= 0.30$ and $\eta_h = 0.40$ herein. Out-of-plane thickness $t$ with plane-stress and the objective scaling factor $\lambda_\mathrm{s}$ are same as that used for the compliant crimper mechanism problem. The filter radius is set to $2\times \min(\frac{L_\mathrm{x}}{N_\mathrm{ex}},\,\frac{L_\mathrm{y}}{N_\mathrm{ey}})$. Other design parameters are equal to those mentioned in Table \ref{Table:T1}.

The symmetric half solution is obtained after $200$ MMA iterations wherein the scaled objective $f_0^\mathrm{CM}= -369.69$ is recorded. The output deformation in the desired direction is noted to $\Delta =\SI{0.221}{mm}$. The full optimized continuum and solution with the pressure field are depicted in Fig.~\ref{fig:sec4fig7b} and Fig.~\ref{fig:sec4fig7c}, respectively. The convergence history plot with some intermediate solutions is shown in Fig.~ \ref{fig:sec4fig7d}. Again some thin sections/flexures (Fig.~\ref{fig:sec4fig7b}) are observed in the optimized design, which help achieve the desired displacement at the output point. Fig.~\ref{fig:sec4fig10b} depicts the deformed profile of the compliant inverter mechanism.

Following the previous research articles, e.g., \cite{frecker1997topological,deepak2009comparative,wang2011projection,vasista2012design} and references therein, to design the compliant crimper and inverter mechanisms, the available symmetric conditions have been employed. However, note that if these symmetric conditions are not used, the optimum results may be different than those presented in Fig.~\ref{fig:sec4fig5b}  and Fig.~\ref{fig:sec4fig7b} due to mesh effects, numerical noise, etc.
\begin{figure*}[h!]
	\begin{subfigure}[t]{0.465\textwidth}
		\centering
		\includegraphics[scale=0.80]{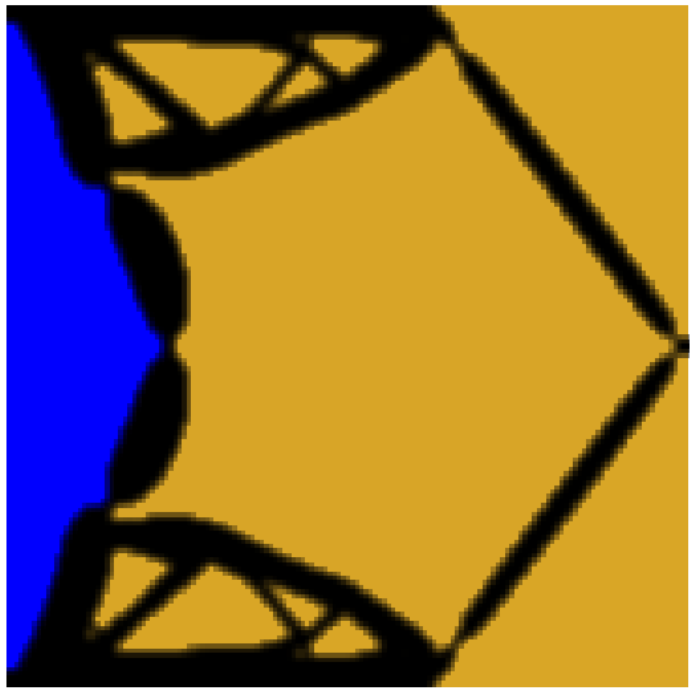}
		\caption{$f_0^\mathrm{CM} = {-1038.89},\,\mathrm{\Delta} = \SI{0.579}{\milli\meter}$}
		\label{fig:sec4fig8a}
	\end{subfigure}
	\begin{subfigure}[t]{0.465\textwidth}
		\centering
		\includegraphics[scale=0.80]{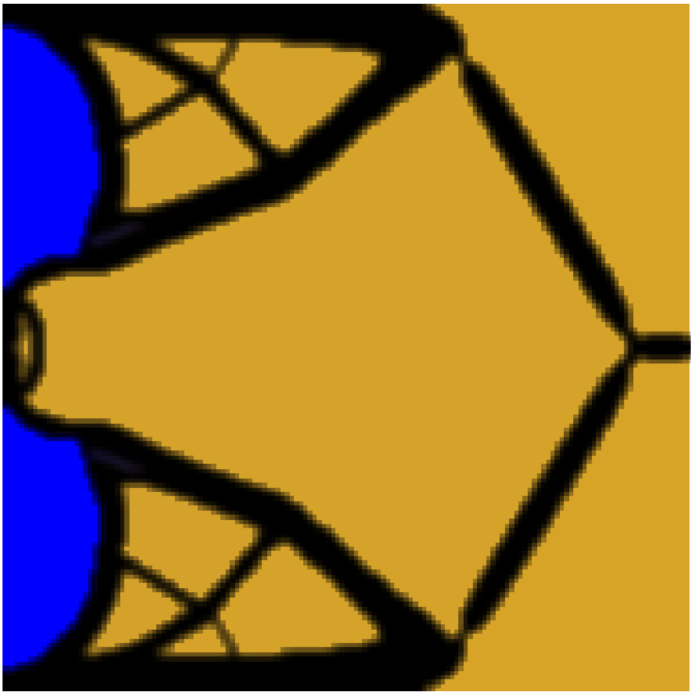}
		\caption{$f_0^\mathrm{CM} = {-253.35},\,\mathrm{\Delta} = \SI{0.162}{\milli\meter}$}
		\label{fig:sec4fig8b}
	\end{subfigure}
	\begin{subfigure}[t]{0.465\textwidth}
		\centering
		\includegraphics[scale=0.8]{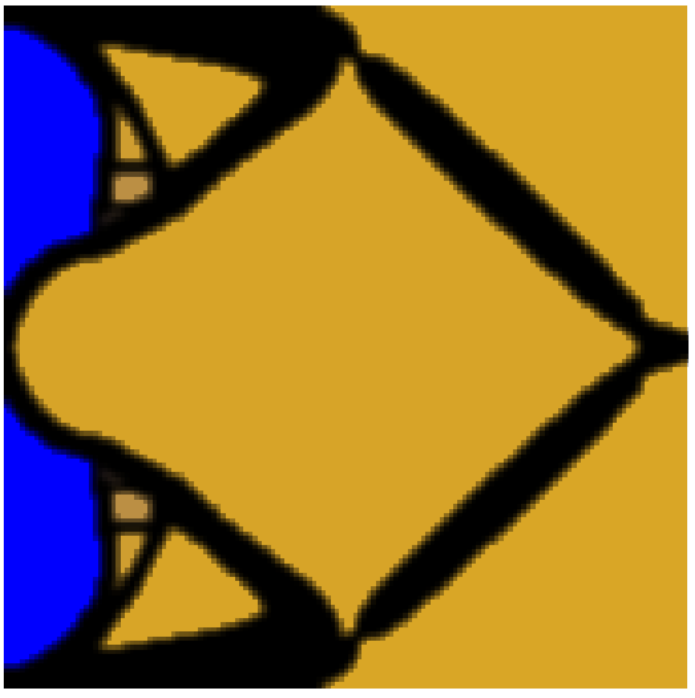}
		\caption{$f_0^\mathrm{CM} = {-55.715},\,\mathrm{\Delta} = \SI{0.0397}{\milli\meter}$}
		\label{fig:sec4fig8c}
	\end{subfigure}
	\begin{subfigure}[t]{0.465\textwidth}
		\centering
		\includegraphics[scale=1]{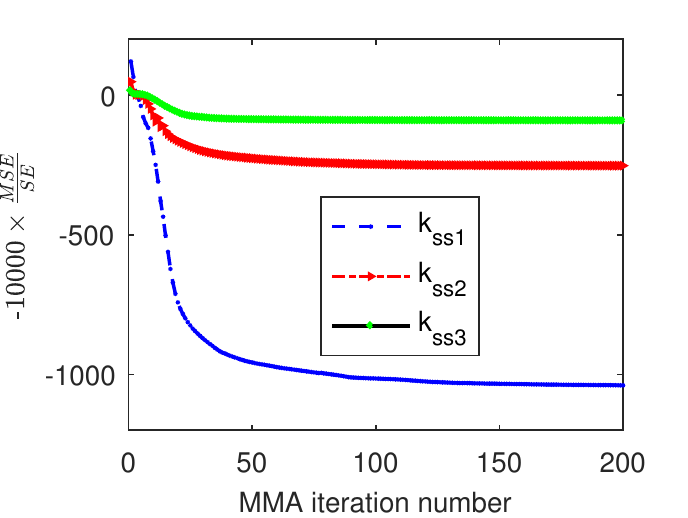}
		\caption{Convergence history}
		\label{fig:sec4fig8d}
	\end{subfigure}
	\caption{Solution to pressure actuated inverter mechanism problem with different output spring stiffnesses. (\subref{fig:sec4fig8a}) Optimized inverter mechanism with spring stiffness $k_\mathrm{ss1}= \SI{5e3}{\newton\per\meter}$ (\subref{fig:sec4fig8b}) Optimized inverter mechanism with spring stiffness $k_\mathrm{ss2}= \SI{1e5}{\newton\per\meter}$ (\subref{fig:sec4fig8c}) Optimized inverter mechanism with spring stiffness $k_\mathrm{ss3}= \SI{1e6}{\newton\per\meter}$ (\subref{fig:sec4fig8d}) Convergence history plot.}\label{fig:sec4fig8}
\end{figure*}
\subsection{Solutions without load sensitivities}\label{LS}
In this section, we demonstrate the effect of the load sensitivities (Eq.~\ref{sec3:eq8} and Eq.~\ref{sec3:eq10}) for designing the pressure loaded piston (Fig.~\ref{fig:sec4fig3}) and pressure actuated compliant crimper mechanism (Fig.~\ref{fig:sec4fig5}). Fig.~\ref{fig:sec4fig11a} and Fig.~\ref{fig:sec4fig11c}  show their optimized continua without using respective load sensitivities (LS).  One notices that the obtained continua in Fig.~\ref{fig:sec4fig11a} and Fig.~\ref{fig:sec4fig11c} are different than those obtained with the full sensitivities shown in Fig.~\ref{fig:sec4fig3c} and Fig.~\ref{fig:sec4fig5c} respectively. In addition, Fig.~\ref{fig:sec4fig11b} and Fig.~\ref{fig:sec4fig11d} depict the magnitude of the LS for the compliance (Eq.~\ref{sec3:eq2}) and the multi-critria (Eq.~\ref{sec3:eq3}) objectives  respectively. One can note, though the magnitude of the LS for the former objective is negligible (Fig.~\ref{fig:sec4fig11b}), it does have influence on the final optimized piston design (Fig.~\ref{fig:sec4fig11a}). In case of pressure actuated CM designs, the magnitude of the LS is comparable to that of the multi-criteria objective (Fig.~\ref{fig:sec4fig11d}) and hence, cannot be neglected. Therefore, considering LS is essential while designing pressure loaded design problems, in particular for compliant mechanisms, and the approach presented herein facilitates easy and computationally inexpensive implementation of the LS within a topology optimization setting.

\subsection{Parameter Study}
The section presents the effect of the different parameters on the obtained designs in several of the aforementioned pressure loaded design problems.
\subsubsection{Volume Fraction}
Herein, a sweep of different volume fractions is performed using the internally pressurized arch-structure problem (Fig.~\ref{fig:sec4fig1a}). It is well known in TO  that  different permitted volume fractions can yield different results \citep{Bendsoe2003}.  

Solutions with  volume fractions $0.075,\,0.1$ and $0.45$, i.e. both lower and higher values compared to Section \ref{IPAS}, are shown in Fig.~ \ref{fig:sec4fig2a},  Fig.~\ref{fig:sec4fig2b} and   Fig.~\ref{fig:sec4fig2c}, respectively. These figures also depict the associated pressure fields. The  convergence history plot for the three cases is illustrated via Fig.~\ref{fig:sec4fig2d}. Evidently, the respective compliance increases with increase in the volume fraction (Figs. \ref{fig:sec4fig2a}$-$\ref{fig:sec4fig2c}). Note that still good results are obtained for fairly low volume fractions. A lower volume fraction may be essential while designing soft structures, single layer, and inflated kind of designs. The present method can be used with suitable boundary conditions for such design problems. 

\subsubsection{Flow resistance and drainage parameters}

The pressure loaded piston design problem is chosen to illustrate the effect of different interpolation parameters, e.g., $\beta_\mathrm{h}$, $\beta_\mathrm{k}$, $\eta_\mathrm{h}$ and $\eta_\mathrm{k}$ on the final solution. Volume fraction $V^*=0.25$ and filter radius $r_\mathrm{min} = 1.8\times\min({\frac{L_\mathrm{x}}{N_\mathrm{ex}},\,\frac{L_\mathrm{y}}{N_\mathrm{ey}}})$ are taken. Note, $\beta_\mathrm{h}$ and $\beta_\mathrm{k}$ control the slopes of $ K(\rho)-\rho$ and $H(\rho)-\rho$ (Figs. \ref{fig:sec2fig1} and \ref{fig:sec2fig3}) plots, respectively. For higher $\beta_\mathrm{k}$, the FEs with $\rho\ge\eta_\mathrm{k}$ behave as solid. Likewise, at high $\beta_\mathrm{h}$, the drainage coefficient of the FEs with $\rho>\eta_\mathrm{h}$ is $h_\mathrm{s}$ (solid elements). In elements where $H(\rho) = 0$, drainage will not be effective indicating void elements.

Fig.~\ref{fig:sec4fig6} shows the optimized continua with respective pressure field for different $\beta$ and $\eta$  after 100 MMA iterations, where all designs had stabilized. While there are global similarities between all designs, it can be noticed that the structural details generated by the proposed method depend on the $\beta$ and $\eta$ parameters. In addition, one also notices that leaking of the inner boundary occurs in Figs. \ref{fig:sec4fig6a}, \ref{fig:sec4fig6c}, \ref{fig:sec4fig6d} and \ref{fig:sec4fig6e}.  This leaking is enabled by a narrow  pathway, from the pressurized domain to the holes in the structure, as seen in the figures. It does not have a significant effect on performance, and this may be the reason why the optimization process does not seem to counteract this tendency. By increasing $\beta$ and decreasing $\eta$, porous boundary regions are smaller which helps to prevent leaks. This is the case in Fig.~\ref{fig:sec4fig6f}, which however also has the worst compliance value. More moderate parameter settings result in a smoother optimization problem and better performance, but in this case with an possibility for  further fluid penetration into the structure. The results still easily permit interpretation as leaktight designs. In general, while choosing $\beta$ and $\eta$ one needs a suitable trade-off between differentiability and decisiveness in defining the boundary.  By and large, as per our experience, $\eta$ close to the volume fraction and $\beta$ in the range of $10$-$20$ provide the required trade-off. 

\subsubsection{Output Spring Stiffness}
As aforementioned, the output spring stiffness drives the TO algorithm to ensure a material connection between the output port and the actuation location. Here, a study with three different spring stiffnesses is presented on the pressure-actuated inverter mechanism problem. 

 Fig.~\ref{fig:sec4fig8a}, Fig.~\ref{fig:sec4fig8b} and Fig.~\ref{fig:sec4fig8c} depict the solution to compliant inverter mechanism problem with $k_\mathrm{ss1}= \SI{5e3}{\newton\per\meter}$, $k_\mathrm{ss2}= \SI{1e5}{\newton\per\meter}, \text{and}\, k_\mathrm{ss3}= \SI{5e5}{\newton\per\meter}$ spring stiffness, respectively. The solutions obtained from symmetric half design are suitably transformed into their respective full continua. The pressure field is also shown for each solution. As expected, as the spring stiffness increases the output deformation decreases. In addition, comparatively more distributed compliance members of the mechanism are obtained for higher output stiffness, and fewer low-stiffness flexures. Note that spring with significantly large $k_\mathrm{ss}$ would give stiff structures. One notices that as spring stiffness increases, area of penetration of pressure within the design domain decreases, i.e., stiffness of the mechanisms increase. With increase in spring stiffness, the corresponding final objective value increases. It has been observed before, that the use of different spring stiffnesses at output port yield different topologies for regular compliant mechanisms problem \citep{deepak2009comparative}. For pressure-actuated compliant mechanisms, one can notice the same trend,  with the lower-stiffness design (Fig.~\ref{fig:sec4fig8}) exploiting a fundamentally different mechanism solution compared to the higher-stiffness cases. The convergence history plots with different spring stiffnesses are shown in Fig.~\ref{fig:sec4fig8d}. 
	\section{Conclusions}\label{Sec5}
In this paper, a novel approach to perform topology optimization of design problems involving both \textit{pressure loaded structures} and \textit{pressure-actuated compliant mechanisms} is presented in a density-based setting. The approach permits use of standard finite element formulation and does not require explicit boudary description or tracking. 

As pressure loads vary with the shape and location of the exposed structural boundary, a main challenge in such problems is to determine design dependent pressure field and its design sensitivity. In the proposed method, Darcy's law in conjunction with a drainage term is used to define the design dependent pressure field by solving an associated PDE using the standard finite element method. The porosity of each FE is related to its material density via a smooth Heaviside function to ensure a smooth transition between void and solid elements. The drainage coefficient is also related to material density using a similar Heaviside function. The determined pressure field is further used to find the consistent nodal loads. In the early stage of the optimization, the obtained nodal loads are spread out within the design domain and thus, may enhance exploratory characteristics of the formulation and thereby the ability of the optimization process to find well-performing solutions. 

The Darcy's parameters, selected \textit{a priori} to the optimization, affect the topologies of the final continua, and recommended values are provided based on the reported numerical experiments. The method facilitates analytical calculation of the load sensitivities  with respect to the design variables using the computationally inexpensive adjoint-variable method. This availability of load sensitivities is an important advantage over various earlier approaches to handle pressure loads in topology optimization. In addition, it is noticed that consideration of load sensitivities within the approach does alter the final optimum designs, and that the load sensitivity terms are particularly important when designing compliant mechanisms. Moreover, in contrast to methods that use explicit boundary tracking, the proposed Darcy method offers the potential for relatively straightforward extension to 3D problems.

The effectiveness and robustness of the proposed method is verified by minimizing compliance and multi-criteria objectives for designing pressure-loaded structures and compliant mechanisms, respectively with given resource constraints. The method allows relocation of the pressure-loaded boundary during optimization, and smooth and steady convergence is observed. Extension to 3D structures and large displacement problems are prime directions for future research.

\section*{Acknowledgment}
The authors are grateful to Krister Svanberg for providing the MATLAB implementation of his Method of Moving Asymptotes, which is used in this work.
   
	\begin{appendices}
		\numberwithin{equation}{section}
\section{Relationship between drainage and penetration depth}\label{appendA}
The ordinary differential equation (ODE) for 1D flow problem using the Darcy flow model with a drainage term can be written as:
\begin{equation}\label{append:eq1}
K(\rho_e)\der{\,^2p}{s^2} = pH(\rho_e),
\end{equation}
where $K$, $p$, and $H$ are the flow coefficient, the pressure and the drainage coefficient, respectively. Since the behavior of pressure  field is simulated that penetrates the material, $\rho_e =1$ is taken for the solution of Eq.~\eqref{append:eq1}. Now, in view of  Eqs. \eqref{sec2:eq2} and \eqref{sec2:eq4}, Eq.~\eqref{append:eq1} can be written as:
\begin{equation}\label{append:eq2}
k_\mathrm{s}\der{\,^2p}{s^2} = ph_\mathrm{s}.
\end{equation}
The motive herein is to express $h_\mathrm{s}$ in terms of the parameters like penetration depth $\Delta s$, the ratio $r$ of the input pressure $p_\mathrm{in}$ and $k_\mathrm{s}$. The following boundary conditions are considered :
\begin{equation}\label{append:eq3}
\begin{rcases}
(i)& \,\lim_{s\to\infty} p = p_\mathrm{out} = 0 \\
(ii)&\, p|_{(s=0)} = p_\mathrm{in}
\end{rcases}.
\end{equation}

A trial solution of Eq.~\eqref{append:eq2} can be chosen as:
\begin{equation}\label{append:eq4}
p(s) = ae^{-bs} + ce^{bs},
\end{equation}
where $e$ is Euler's number and $a,\,b,\,\text{and}\, c$ are unknown coefficients which are determined using the above boundary conditions as:
\begin{equation}\label{append:eq5}
a = p_\mathrm{in},\, b = \sqrt{\frac{h_\mathrm{s}}{k_\mathrm{s}}},\,c = 0.
\end{equation}

\noindent Thus,
\begin{equation}\label{append:eq6}
p(s) = p_\mathrm{in}e^{-\sqrt{\frac{h_\mathrm{s}}{k_\mathrm{s}}}s}
\end{equation}

With $p|_{(s = \Delta s)}= rp_\mathrm{in}$, Eq.~\eqref{append:eq6} yields:

\begin{equation}
h_\mathrm{s} =\left(\frac{\ln{r}}{\Delta s}\right)^2 k_\mathrm{s}.
\end{equation}
		\section{Evaluating the Lagrange Multipliers}\label{appendA2}
\numberwithin{equation}{section}
Here, the calculation procedure for the Lagrange multipliers $\bm{\lambda}_1,\,\bm{\lambda}_2$ and $\bm{\lambda}_3$ is presented. To clarify the process, we partition the displacement and pressure vectors. Say, subscripts $\mathrm{u}$ and $0$ indicate the free and prescribed degrees of freedom for the displacement vector $\mathbf{u}$, and subscripts $\mathrm{f}$ and $\mathrm{p}$ denote the free and prescribed degrees of freedom for the pressure vector $\mathbf{p}$. Therefore,
\begin{equation} \label{eq:appe2_1}
\mathbf{u} = \begin{bmatrix}
\mathbf{u}_\mathrm{u}\\
 \mathbf{u}_0
\end{bmatrix},\qquad\mathbf{p} = \begin{bmatrix}
\mathbf{p}_\mathrm{f}\\
\mathbf{p}_\mathrm{p}
\end{bmatrix}.
\end{equation}
Likewise, the global stiffness matrix $\mathbf{K}$, the global conversion matrix $\mathbf{H}$ and the the global 
flow matrix $\mathbf{A}$ can also be partitioned as:
\begin{equation}\label{eq:appe2_2}
\mathbf{K} = 
\begin{bmatrix}
\mathbf{K}_{\mathrm{uu}} & \mathbf{K}_{\mathrm{u0}} \\
\mathbf{K}_{\mathrm{0u}} & \mathbf{K}_{\mathrm{00}} \\
\end{bmatrix},\, \mathbf{H} = 
\begin{bmatrix}
\mathbf{H}_{\mathrm{uf}} & \mathbf{H}_{\mathrm{up}} \\
\mathbf{H}_{\mathrm{pu}} & \mathbf{H}_{\mathrm{0p}} \\
\end{bmatrix},\, \mathbf{A} = \begin{bmatrix}
\mathbf{A}_{\mathrm{ff}} & \mathbf{A}_{\mathrm{fp}} \\
\mathbf{A}_{\mathrm{pf}} & \mathbf{A}_{\mathrm{pp}} \\
\end{bmatrix}.
\end{equation}
Note that the derivatives $\frac{\partial \mathbf{u}_0}{\partial\bm{\rho}} = 0$ and $\frac{\partial \mathbf{p}_\mathrm{p}}{\partial\bm{\rho}} = 0$ as $\mathbf{u}_0 $ and $ \mathbf{p}_\mathrm{p}$ are prescribed and they do not depend upon the design vector. Now, using these facts with the partitioned descriptions of matrices (Eq.~\ref{eq:appe2_2}),  Eq.~\ref{sec3:eq5} can be rewritten as
  \begin{equation}\label{eq:appe2_3}
\begin{aligned}
\frac{d\mathrm{\Phi}}{d\bm{\rho}} =&
\underbrace{\left(\pd{f_0}{\mathbf{u}_\mathrm{u}} + \tr{\bm{\lambda}_1^\mathrm{u}} \mathbf{K}_{\mathrm{uu}}\right)}_{\text{Term}\,1}\pd{\mathbf{u}_\mathrm{u}}{\bm{\rho}} + \pd{f_0}{\bm{\rho}} + \tr{\bm{\lambda}}_1\pd{\mathbf{K}}{\bm{\rho}}\mathbf{u}
\\+ &\underbrace{\left( \tr{\bm{\lambda}_1^\mathrm{u}} \mathbf{H}_{\mathrm{uf}} + \tr{\bm{\lambda}_2^\mathrm{f}} \mathbf{A}_{\mathrm{ff}}\right) }_{\text{{Term}} \,2}\pd{\mathbf{p}_\mathrm{f}}{\bm{\rho}}+\tr{\bm{\lambda}}_2\pd{\mathbf{A}}{\bm{\rho}}\mathbf{p} \\
+&\underbrace{\left(\pd{f_0}{\mathbf{v}_\mathrm{u}} + \tr{\bm{\lambda}_3^\mathrm{u}} \mathbf{K}_{\mathrm{uu}}\right)}_{\text{Term}\,3}\pd{\mathbf{v}_\mathrm{u}}{\bm{\rho}} + \tr{\bm{\lambda}}_3\pd{\mathbf{K}}{\bm{\rho}}\mathbf{v},
\end{aligned}
\end{equation} 
where $\bm{\lambda}_1^\mathrm{u},\,\bm{\lambda}_2^\mathrm{f}$ and $\bm{\lambda}_3^\mathrm{u}$ are the Lagrange multiplier vectors for free degrees of freedom corresponding to $\bm{\lambda}_1,\,\bm{\lambda}_2$ and $\bm{\lambda_3}$ respectively, which are selected such that Term 1, Term 2 and Term 3 in Eq.~\eqref{eq:appe2_3} vanish, i.e.,
\begin{equation}\label{eq:appe2_4}
\begin{rcases}
\tr{\bm{\lambda}_1^\mathrm{u}}  &= -\pd{f_0(\mathbf{u},\, \mathbf{v},\,\bm{\rho})}{\mathbf{u}_\mathrm{u}} \inv{\mathbf{K}}_{\mathrm{uu}}\\
\tr{\bm{\lambda}_2^\mathrm{f}}  & = -\tr{\bm{\lambda}_1^\mathrm{u}} \mathbf{H}_{\mathrm{uf}}\inv{\mathbf{A}}_{\mathrm{ff}}\\
\tr{\bm{\lambda}_3^\mathrm{u}}  &= -\pd{f_0(\mathbf{u},\, \mathbf{v},\,\bm{\rho})}{\mathbf{v}_\mathrm{u}} \inv{\mathbf{K}}_{\mathrm{uu}}
\end{rcases}.
\end{equation} 	
The prescribed degrees of freedom of all multipliers are zero, thus Eq.~\eqref{sec3:eq7} holds without partitioning.

	\end{appendices}
	\bibliography{myreference}
	\bibliographystyle{spbasic} 
\end{document}